\documentclass[journal,twoside, 10pt]{IEEEtran}

\usepackage{mathrsfs}
\usepackage{graphicx}
\usepackage{booktabs}
\usepackage{multirow}
\usepackage{multicol}
\usepackage{amsmath}
\usepackage[noend]{algpseudocode}
\usepackage{algorithm}
\usepackage{algorithmicx}
\usepackage{float}
\usepackage{amssymb}
\usepackage{cite}
\usepackage{bm}
\usepackage{textcomp}
\usepackage{subfig}
\usepackage{array, booktabs}
\usepackage{stfloats}
\usepackage{diagbox}
\usepackage{xurl}
\usepackage{makecell}
\usepackage[table,xcdraw]{xcolor}
\usepackage[flushleft]{threeparttable}
\usepackage{fancyhdr}
\usepackage[justification=justified,singlelinecheck=false]{caption}
\usepackage{etoolbox}

\begin{document}

\title{Ultrawideband Polarization-Rotating Antenna for Enhanced Ground Penetrating Radar Detection}

\author{Ming~Li,~\IEEEmembership{Member,~IEEE}, Yuki~Gao,~\IEEEmembership{Graduate~Student~Member,~IEEE}, Shu-Lin~Chen,~\IEEEmembership{Senior~Member,~IEEE}, Hai-Han Sun,~\IEEEmembership{Senior~Member,~IEEE}, and Y.~Jay~Guo,~\IEEEmembership{Life~Fellow,~IEEE}
\thanks{This work was supported in part by the Australian Research Council (ARC) Discovery Early Career Researcher Award (DECRA) under Grant DE250100427, in part by the UW--Madison ECE Distinguished Graduate Fellowship, and in part by the Wisconsin Alumni Research Foundation (WARF). (\textit{Ming Li and Yuki Gao contributed equally to this work}.) (\textit{Corresponding author: Shu-Lin Chen, Hai-Han Sun}.) }
\thanks{Ming Li, Shu-Lin Chen, and Y. Jay Guo are with the Global Big Data Technologies Centre (GBDTC), University of Technology Sydney (UTS), NSW 2007, Australia (email: ming.li@uts.edu.au; shulin.chen@uts.edu.au; jay.guo@uts.edu.au).}
\thanks{Yuki Gao and Hai-Han Sun are with Department of Electrical and Computer Engineering, University of Wisconsin-Madison, Madison, WI 53706 USA (email: gao369@wisc.edu; haihan.sun@wisc.edu).} 
}

\maketitle

\markboth{}{}

\begin{abstract}

Ground penetrating radar (GPR) represents a powerful and widely adopted tool for nondestructive subsurface detection and imaging. 
To overcome the impaired detection performance caused by polarization mismatch in conventional single-channel GPR systems, this paper presents an innovative ultrawideband polarization-rotating (UPR) antenna that achieves rotating linear polarizations (LPs) by varying the source frequency for enhanced detection. 
The developed UPR antenna achieves a continuous polarization rotation from $0^\circ$ to $170^\circ$ across 1.2 to 3.2~GHz, as verified by both full-wave simulations and experimental measurements.
The antenna is subsequently integrated into a single-channel stepped-frequency GPR system and evaluated against targets of varying geometries in a sandy environment. 
Compared with a conventional fixed-LP GPR system, the UPR antenna-enabled GPR system delivers markedly more consistent detection performance across all target types, with particularly pronounced enhancements for polarization-sensitive targets. 
By enabling ultrawideband frequency-dependent polarization agility, the proposed GPR system enhances subsurface target detection while maintaining the simplicity and low cost of a single-channel measurement architecture.
\end{abstract}


\section{Introduction}

\IEEEPARstart {G}{round} penetrating radar (GPR) is a widely used nondestructive testing (NDT) tool for subsurface detection and imaging \cite{daniels2004ground, lee2004novel, h_m_jol_ground_2009}.  A GPR system typically consists of a transceiver and one or more antennas. It operates by transmitting electromagnetic waves into the subsurface and recording the backscattered signals from buried objects and interfaces with wave impedance contrasts. It has been applied to a wide range of near-surface sensing tasks in different fields, including underground utility detection \cite{yamaguchi_mapping_2021, jiang_cable-mapping_2019}, civil infrastructure assessment \cite{wang_automatic_2011, hou_review_2022}, root system mapping \cite{liu2017detection, luo2024automatic, luo_tree_2022}, and unexploded ordnance (UXO) detection \cite{chi-chih_chen_ultrawide-bandwidth_2001}.

In GPR systems, antenna polarization plays a critical role in determining the detection performance, particularly for polarization-sensitive targets commonly encountered in subsurface environments. These targets include elongated or anisotropic objects, such as pipes, cables, metallic rods, tree roots, and structural cracks\cite{soldovieri2010sparse,luo2024automatic, liu2017detection,sun_estimating_2022}, whose backscattered responses are strongly dependent on the polarization of the incident electric field relative to the target orientation. In conventional single-channel GPR systems employing a fixed linear polarization (LP), polarization mismatch readily occurs when the target orientation is unknown or varies across the sensing environment. Under such conditions, poor alignment between the incident electric field and the target can substantially reduce the backscattered signal strength, leading to degraded detection reliability or even missed detections\cite{liu2017detection, lualdi_combining_2014}.

To address the issue, polarimetric GPR has been developed and has attracted widespread attention in the past decades. Fully polarimetric GPR systems typically employ antennas with orthogonal LPs, namely horizontal (H) and vertical (V) polarizations, to acquire four polarimetric scattering components  $S_{HH}$, $S_{HV}$, $S_{VH}$, and $S_{VV}$ \cite{sun2021dual,sun2023diameter,feng_subsurface_2015}. These components form the polarimetric scattering matrix, which can be used to improve detection accuracy and extract target information that is not available from single-polarimetric measurements \cite{zhou2022multiparameter, o2002discrimination, o2002effects, dong2023super, sun2023diameter, liu2023improved, liu2019hybrid, sun2021dual}. Fully polarimetric GPR systems have demonstrated enhanced detection and characterization capabilities across a wide range of subsurface targets, including landmines \cite{zhou2022multiparameter}, UXO \cite{o2002discrimination, o2002effects}, subsurface cracks \cite{dong2023super}, reinforcing bars \cite{sun2023diameter, liu2023improved, liu2019hybrid}, and tree roots \cite{sun_estimating_2022}. Despite these advantages, fully polarimetric GPR measurements typically require multiple antennas connected to multiple transceiver channels, which increases hardware complexity, system cost, and integration difficulty. These challenges become more pronounced in advanced array-based GPR platforms designed for high-resolution sensing and imaging \cite{cheng_early_2024, srivastav_highly_2020}.

A low-complexity approach that enables polarization-agile capability for GPR systems using a single transceiver channel is therefore highly desirable. One promising approach is to employ the frequency-dependent polarization-rotating antenna \cite{chen2024analysis, xiong2025multiexcited, tong2026contiunous}, which provides continuously rotating LPs across its operating frequency band. This feature would allow a single transceiver to interrogate subsurface targets from various LP orientations through frequency sweeping alone, eliminating the need for additional RF channels or mechanical antenna rotation. 
However, despite this potential, such polarization-rotating antennas have not yet been incorporated into GPR measurement systems. Existing designs generally lack the ultrawideband performance required for GPR applications, and the measurement principles, system integration, and experimental validation for subsurface sensing remain largely unexplored.

In this paper, we develop a single-port GPR measurement system incorporating an ultrawideband polarization-rotating (UPR) antenna to achieve polarization-agile subsurface interrogation using a single transceiver channel. The UPR antenna, consisting of two orthogonally arranged Vivaldi radiators, achieves a continuous and frequency-dependent LP rotation from approximately 0° to 170° as the operating frequency varies from 1.2 to 3.2 GHz, corresponding to a fractional bandwidth of 90.9\%. This is achieved without requiring any mechanical rotation of the antenna. By integrating the presented processing and detection methodology, the system enables robust detection while preserving the simplicity of the single-transceiver architecture. This highlights its strong potential for high-performance, low-complexity, and low-cost subsurface sensing applications. Compared to the state of the art, the main contributions of this work are summarized as follows:
\begin{itemize}

\item At the system level, to the best of the authors' knowledge, this work presents the first single-channel stepped-frequency GPR measurement system exploiting frequency-controlled polarization rotation. Compared with a conventional single-port fixed linearly polarized (SPLP) GPR system, the UPR-based system demonstrates noticeably more consistent and robust detection performance, with particularly pronounced enhancements for elongated targets with unknown orientations. Comprehensive experimental measurements validate the effectiveness of the detection system.

\item At the antenna performance level, the developed UPR antenna achieves a 90.9\% fractional bandwidth, substantially extending the operating bandwidth of existing frequency-dependent polarization-rotating antennas and enabling their practical integration into ultrawideband GPR measurement systems. The antenna combines two orthogonally arranged Vivaldi radiators with a frequency-controlled feeding network to realize continuous linear polarization rotation. Furthermore, the presented antenna design and integration strategy establish a general framework for incorporating frequency-dependent polarization-rotating antennas into GPR systems.
 
\end{itemize}

\section{Ultrawideband Polarization-Rotating Antenna}

\begin{figure}[!t]
	\centering		
	\includegraphics[width=3.5in]{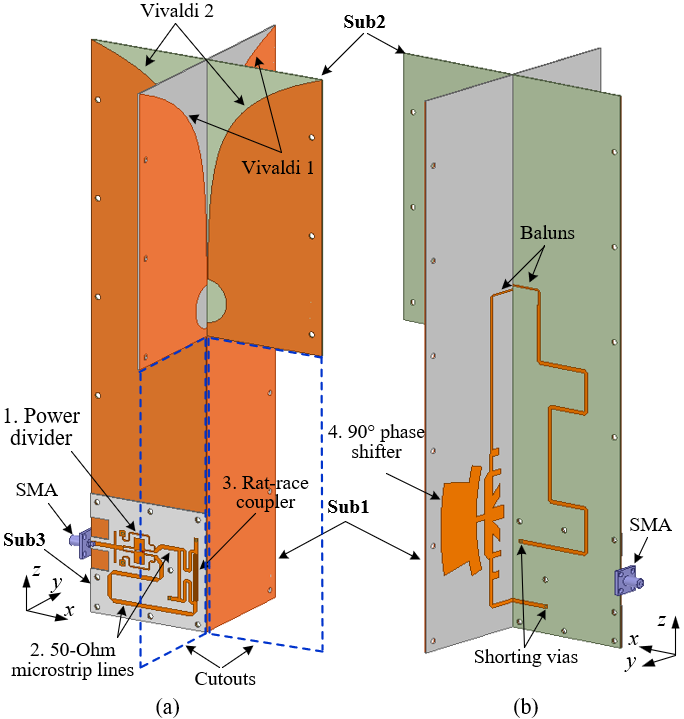}		
	\caption{(a) Front and (b) back view of the designed UPR antenna. Components 1-4 constitute the frequency-controlled feeding network.}
	\label{fig11}
\end{figure}

The configuration of the UPR antenna is shown in Fig.~\ref{fig11}.  
The antenna is constructed using two vertically oriented substrates (Sub1 and Sub2), and they are orthogonally arranged to form a self-supporting cross-shaped structure. Longitudinal slits are introduced along the center of both substrates, with Sub1 slotted from the upper edge and Sub2 from the lower edge, allowing the two substrates to interlock mechanically and provide mutual structural support. To facilitate assembly, rectangular cutouts of 162.5 $\times$ 62.6 mm$^2$ are introduced in both Sub1 and Sub2, as shown in Fig.~\ref{fig11}. In addition, a third substrate (Sub3) is mounted on Sub2 to accommodate the frequency-dependent feeding network. All three substrates are Astra MT77 laminates with metallic patterns printed on both sides \cite{isola_mt77}. They feature a dielectric constant of $\varepsilon_r=$3 and thickness of 0.76~mm. Sub1 and Sub2 feature the same dimensions of 290$\times$126~mm$^2$, whilst Sub3 measures 60$\times$63.4~mm$^2$. \par

\subsection{Ultrawideband Vivaldi Radiators}

Sub1 and Sub2 host two ultrawideband Vivaldi radiators on their top layers, referred to as Vivaldi 1 and Vivaldi 2, as marked in Fig. \ref{fig11}. The two radiators are designed to operate from 1.2 to 3.2~GHz, with identical geometries and feeding structures, except for a slight offset at the cross-over point of the balanced-to-unbalanced (balun) transformer. This offset is introduced to avoid overlap between the feeding lines during assembly while maintaining similar radiation performance. 

\begin{figure}[!t]
	\centering		
	\includegraphics[width=2.8in]{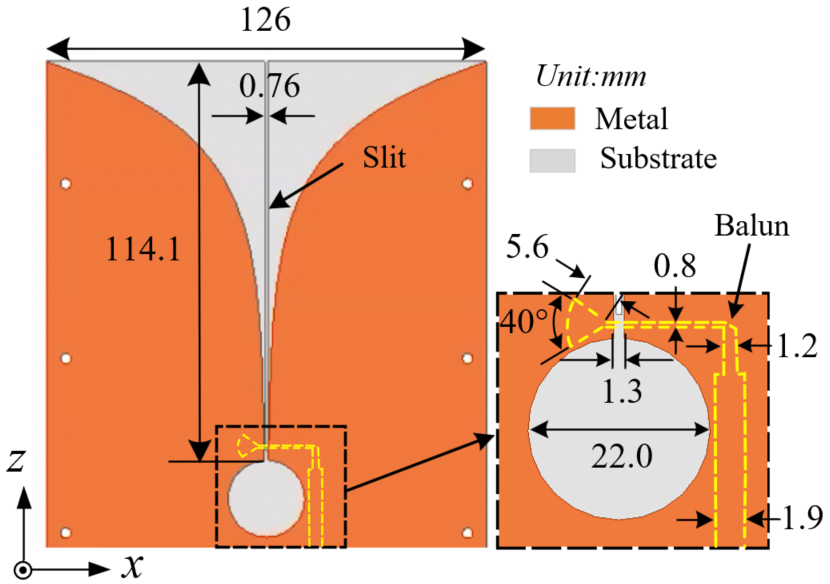}		
	\caption{The Vivaldi radiator 1 on Sub1, which consists of an exponentially flared slot and a circular cavity on the top layer, and a two-stage feeding balun on the bottom layer.}
	\label{fig111}
\end{figure}

A closed view of the Vivaldi radiator is shown in Fig.~\ref{fig111}. Each radiator has an exponentially flared slot that gradually expands from the feed region to the aperture, enabling a smooth transition from guided waves to radiating waves.  To improve impedance matching over an ultrawide operating bandwidth, a circular cut is introduced near the slot origin, with the dimensions detailed in Fig.~\ref{fig111}. Additionally, a two-stage balun transformer is designed on the bottom layer and coupled to the radiator at the slot origin \cite{dzagbletey2019quarter}.  As shown in the inset of Fig.~\ref{fig111}, the balun transforms the 50-$\Omega$ microstrip line to an open-circuit radial stub, facilitating ultrawideband impedance matching with the tapered slot. 
The orthogonally-arranged Vivaldi 1 and Vivaldi 2 are designed to generate electromagnetic waves with orthogonal LPs. These two LP states will be modulated by a specifically designed ultrawideband feeding network to produce frequency-dependent LP rotation.

\subsection{Ultrawideband Frequency-Controlled Feeding Network}

To simultaneously excite Vivaldi 1 and Vivaldi 2 and realize ultrawideband LP rotation, an ultrawideband frequency-controlled feeding network is developed. As illustrated in Fig.~\ref{fig11}, the feeding network comprises four key components: a 3-dB power divider, a pair of 50-$\Omega$ microstrip lines with unequal electrical lengths, a rat-race coupler, and a  $90^\circ$  phase shifter. All components are designed to cover the target ultrawide frequency range of 1.2 to 3.2~GHz. \par

\begin{figure}[!t]
	\centering			
	\includegraphics[width=3.0in]{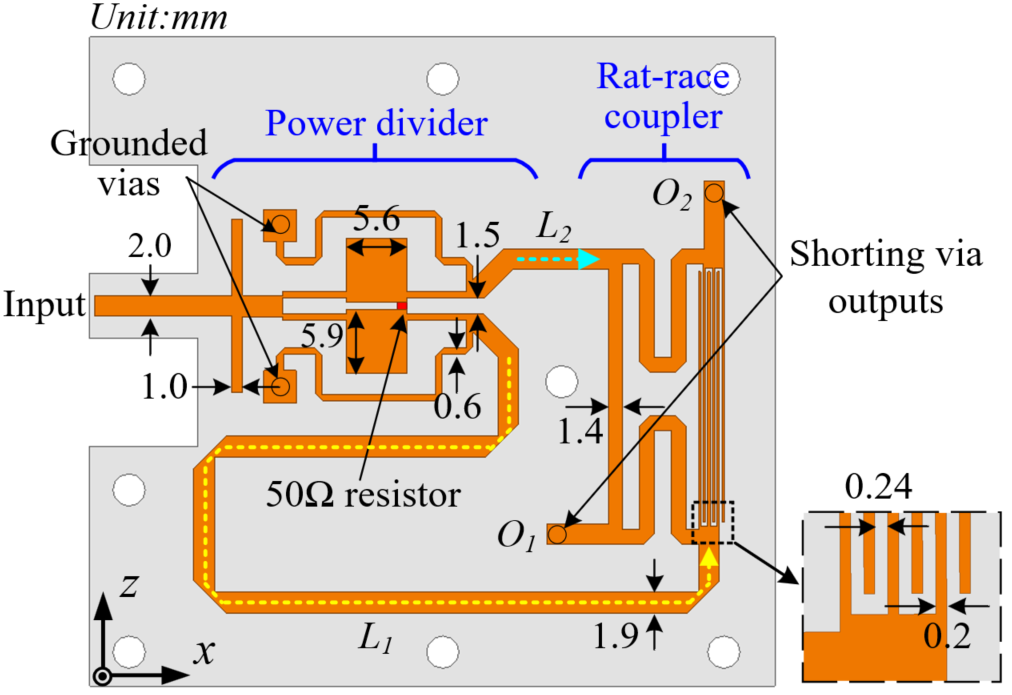}		
	\caption{Wideband power divider and rat-race coupler implemented on Sub3. The two components are connected through a pair of unequal-length 50-$\Omega$ microstrip lines, with $L_1 = 94.5$~mm and $L_2 = 8.9$~mm, to introduce a frequency-controlled phase difference.}
	\label{fig2}
\end{figure}

Fig.~\ref{fig2} shows the detailed configuration of the power divider and the rat-race coupler implemented on Sub3.  The power divider adopts open stubs at the input, and quasi-coupled lines and short stubs at the two outputs to realize a wide operating bandwidth \cite{tang2016design}. The two outputs of the power divider are connected to a pair of 50-$\Omega$ microstrip lines with unequal electrical lengths ($L_1$ and $L_2$), which introduce a frequency-controlled dispersive phase difference of $2\Psi$ between the two signal paths. Specifically, $2\Psi$ is designed to vary linearly from $-180^\circ$ to $180^\circ$ across the operating frequency band. The detailed performance will be discussed in the next Section. The two phase-shifted paths are then connected to the input ports of the rat-race coupler. The rat-race coupler uses three transmission-line sections and a shorted parallel-coupled six-line to provide wideband sum-and-difference signal combination while maintaining a compact size \cite{zheng2020single}. \par

\begin{figure}[!t]
	\centering		
	\includegraphics[width=2.4in]{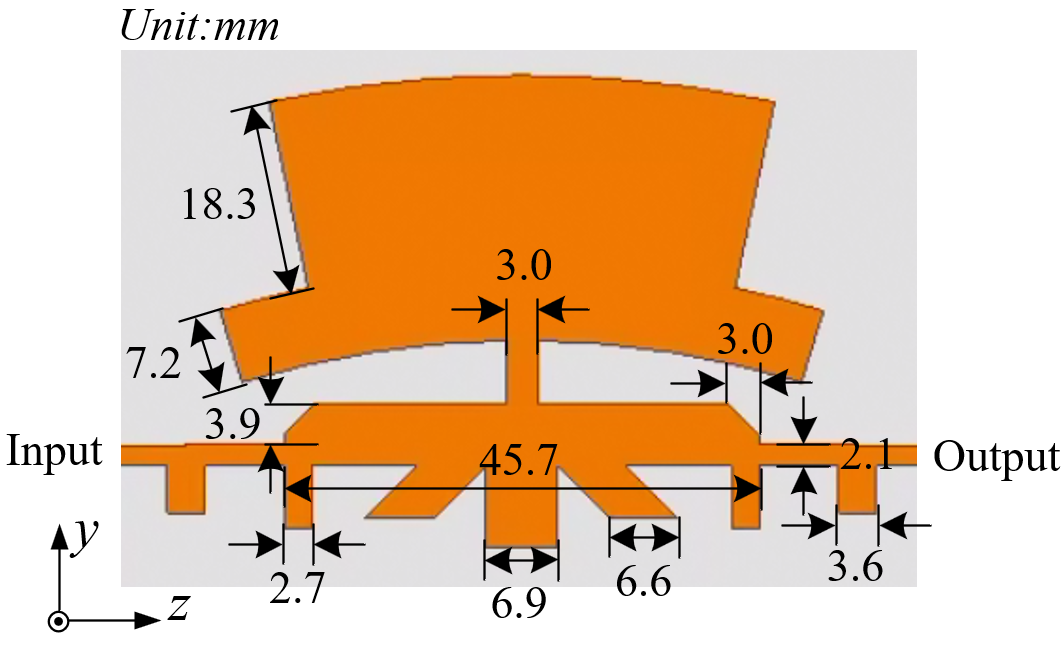}		
	\caption{The wideband $90^\circ$ differential phase shifter. }
	\label{fig4}
\end{figure}

The two output ports ($O_{1}$ and $O_{2}$) of the rat-race coupler on Sub3 are connected to the microstrip lines on the bottom layer of Sub2 through shorting vias isolated from the ground plane, as marked in Fig.~\ref{fig11}(b). One microstrip line is connected to the balun to excite Vivaldi 2 on Sub2, while the other is connected to the microstrip line on Sub1 through a vertical transition at the substrate edges. The microstrip line on Sub1 is then connected to a wideband $90^\circ$ differential phase shifter before exciting Vivaldi 1. The detailed configuration of the  differential phase shifter is shown in Fig.~\ref{fig4} \cite{yeung2013design}. It provides a constant $90^\circ$ phase difference across the operating bandwidth with respect to the reference feeding path on Sub2. 

For the fully assembled UPR antenna, the input signal is fed through an SMA connector mounted on Sub3. Assuming the input voltage has an amplitude of $V$, the total radiated electric field can be expressed as \cite{chen2024analysis}:
\begin{equation}
	\vec{E}_{\text{tot}} \propto V \sin \Psi \hat{e}_x + V \cos \Psi \hat{e}_y = V \hat{e}_t,
	\label{eq3}
\end{equation}
where $\hat{e}_x$ and $\hat{e}_y$ denote unit vectors along the $x$- and $y$-axes, respectively. From (\ref{eq3}), the resulting LP direction $\hat{e}_t$ of the UPR antenna is determined by $\Psi$. By properly designing the lengths of $L_1$ and $L_2$, the phase difference $\Psi$ increases continuously from $-90^\circ$ to $90^\circ$ as the frequency increases over the operating band from 1.2 to 3.2~GHz. Consequently, a continuous $180^\circ$ rotation of the LP direction can be achieved.

\begin{figure}[!t]
		\centering
		\includegraphics[width=2.35in]{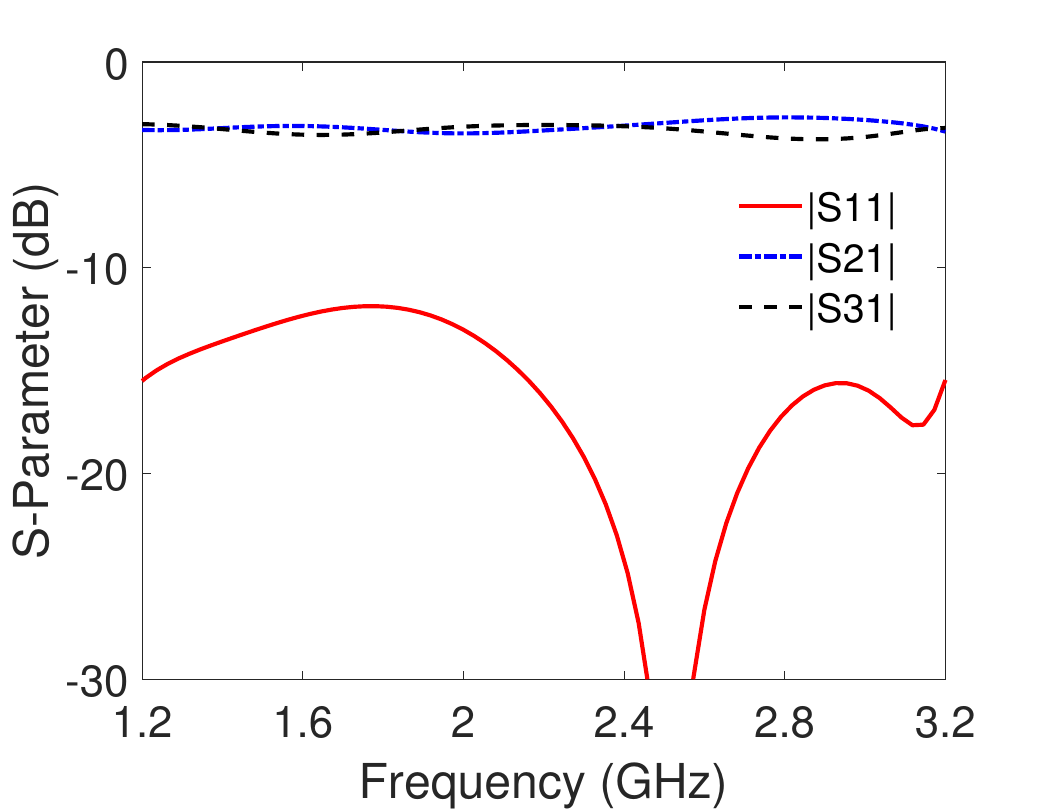}
		\vspace{-0.5em}
		\begin{center} \footnotesize{(a)} \end{center}
	\vspace{-0.5em}
		\centering
		\includegraphics[width=2.4in]{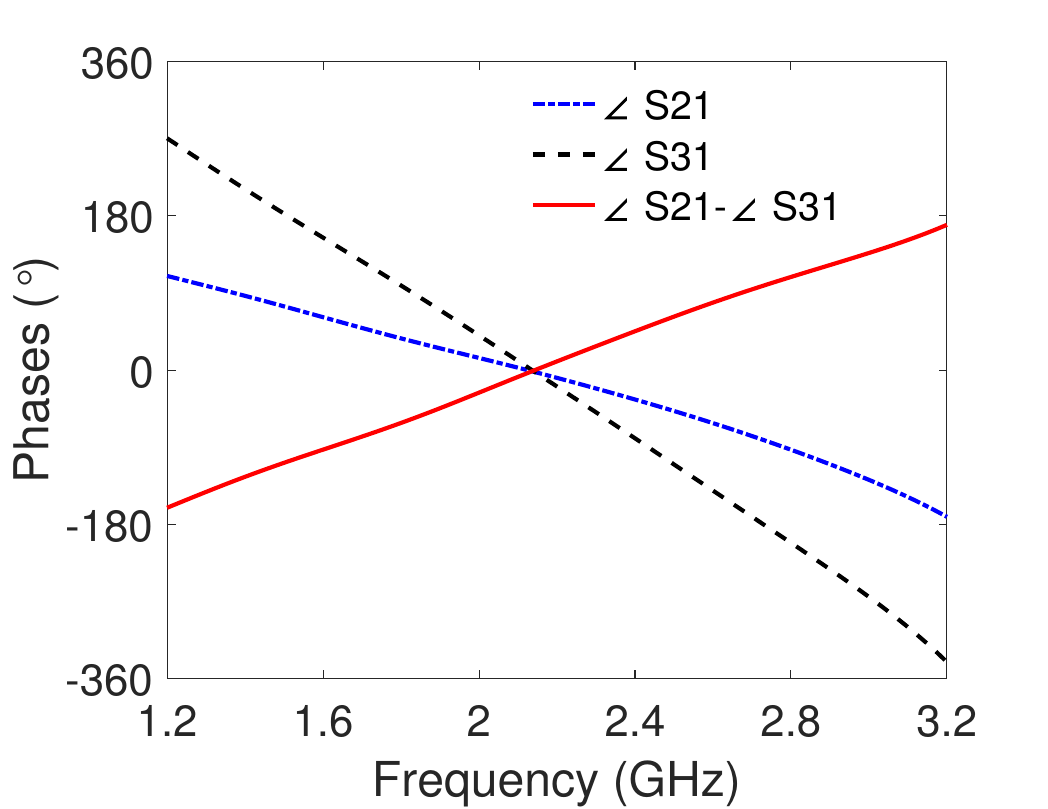}
		\vspace{-0.5em}
		\begin{center} \footnotesize{(b)} \end{center}
	\caption{ (a) Simulated magnitudes and (b) phases of the S-parameters of the wideband power divider connected to a pair of 50-$\Omega$ microstrip lines with unequal lengths. }
	\label{fig200}
\end{figure}

\section{Antenna Simulation and Measurement}

\begin{figure}[t]
	\centering
	\begin{tabular}{cc}
		\includegraphics[width=0.45\linewidth]{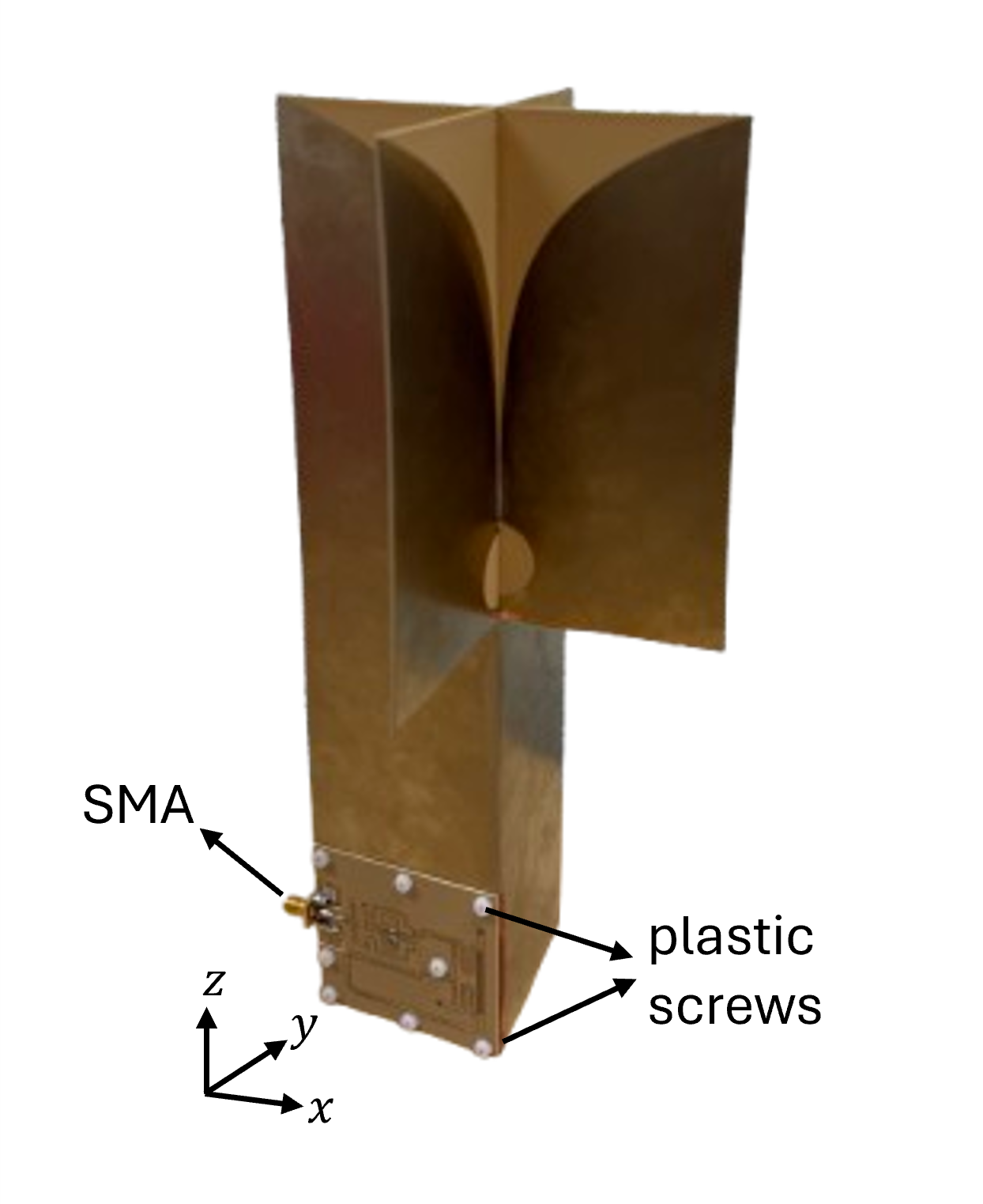} &
		\includegraphics[width=0.45\linewidth]{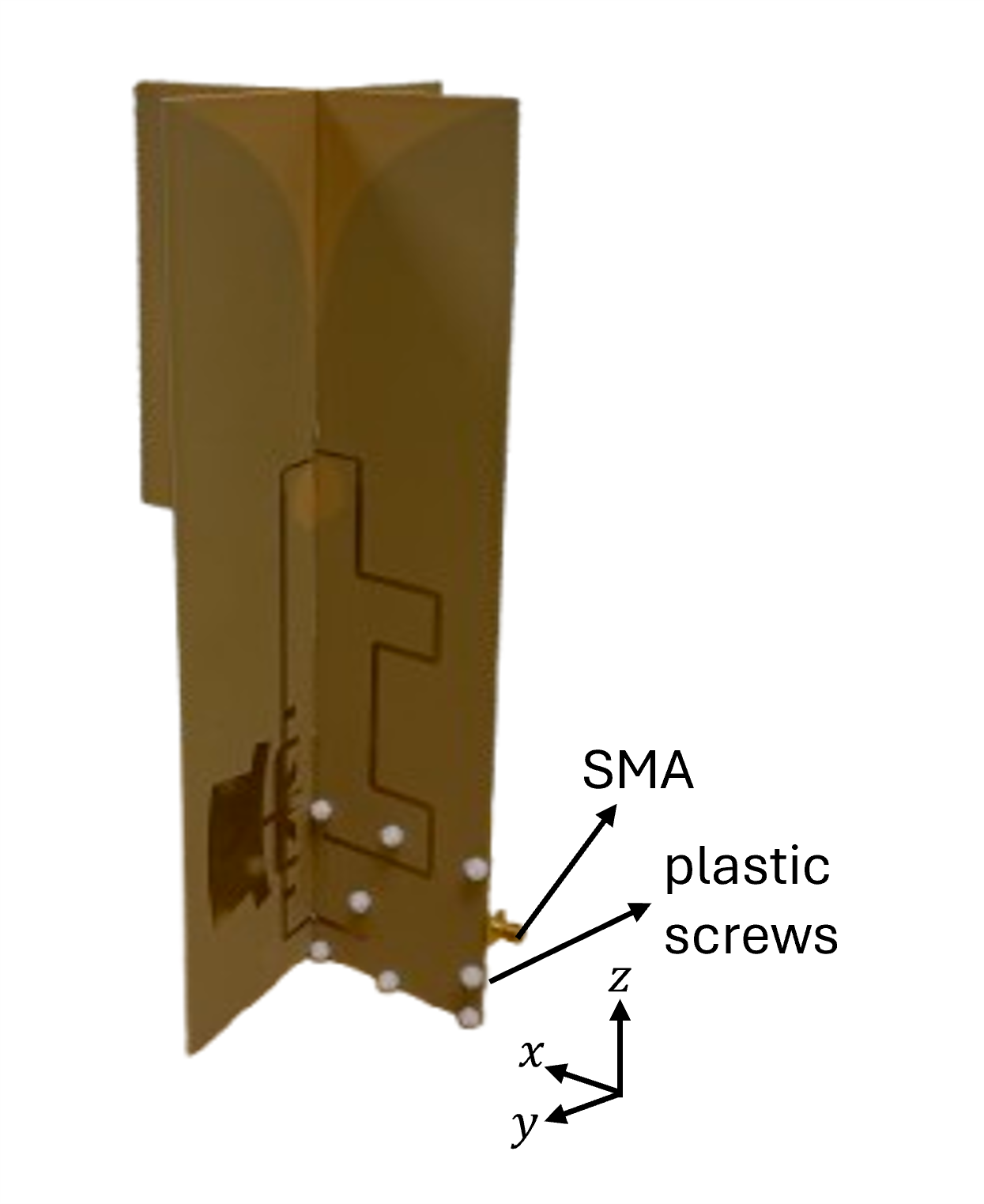} \\
		\footnotesize{(a)} & \footnotesize{(b)}
	\end{tabular}
	\caption{(a) Front and (b) back view of the designed UPR antenna prototype.}
	\label{fig100}
\end{figure}

\begin{figure}[!t]
	\centering		
	\includegraphics[width=2.7in]{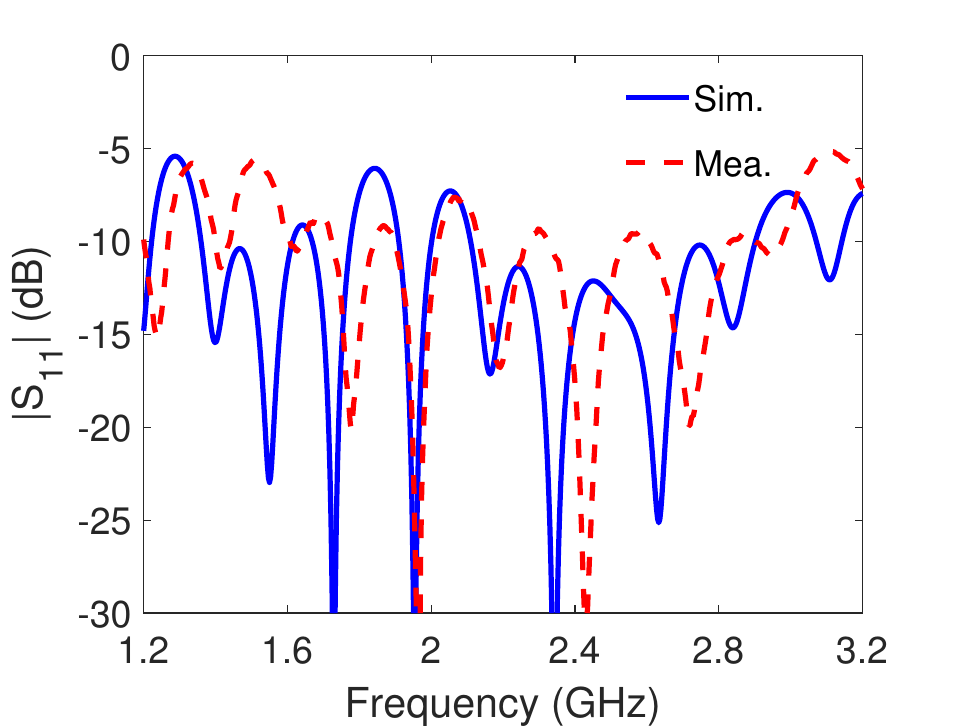}		
	\caption{Simulated and measured $|S_{11}|$ of the UPR antenna.}
	\label{fig101}
\end{figure}

\begin{figure}[!t]
	\centering		
	\includegraphics[width=2.7in]{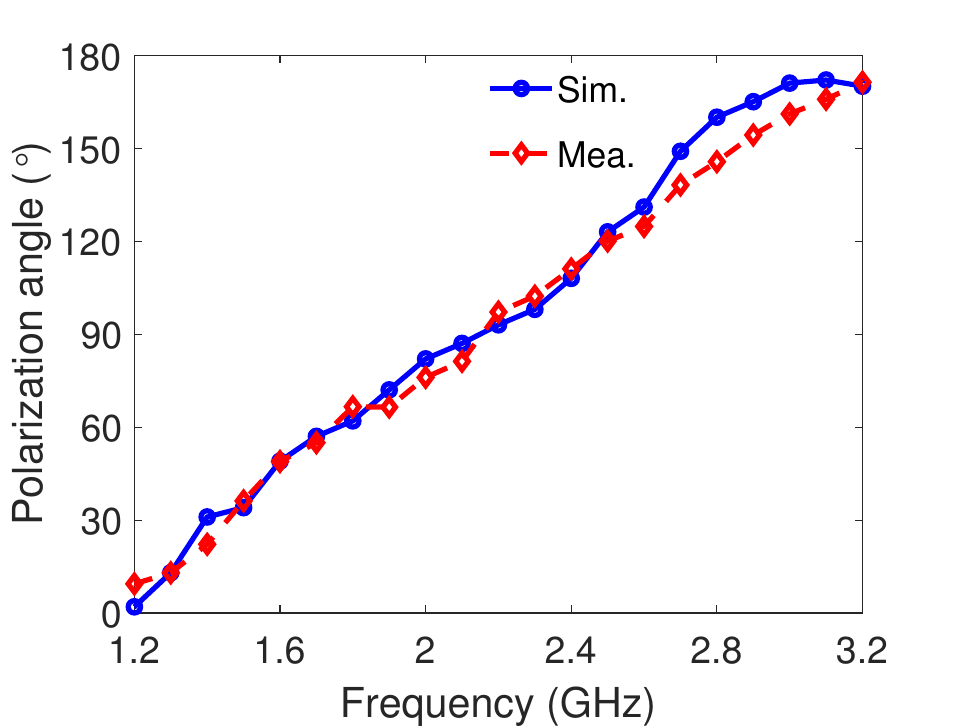}		
	\caption{Simulated and measured LP angle of the developed UPR antenna.}
	\label{fig104}
\end{figure}

\begin{figure}[!t]
	\centering		
	\includegraphics[width=3.1in]{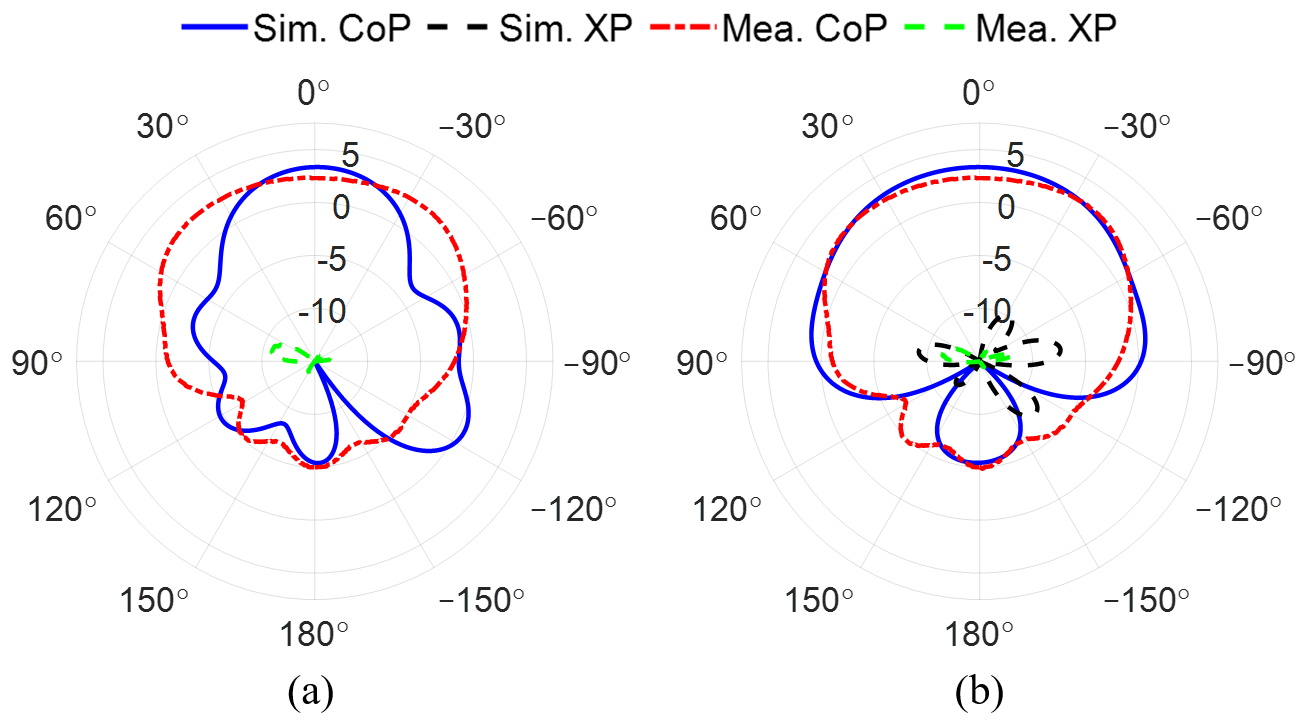}		
	\vspace{0.5em}
	\newline
	\includegraphics[width=3.1in]{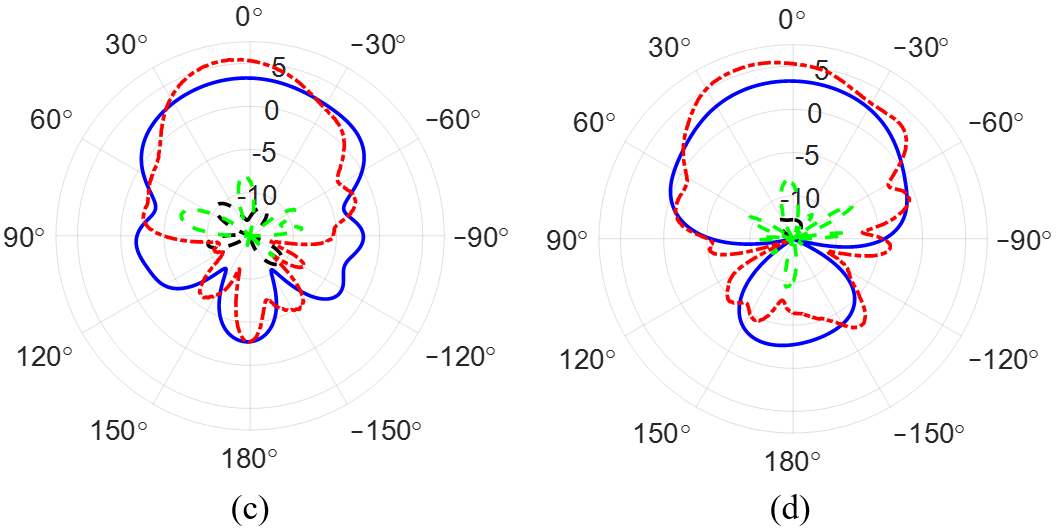}	
	\vspace{0.5em}	
	\newline
	\includegraphics[width=3.1in]{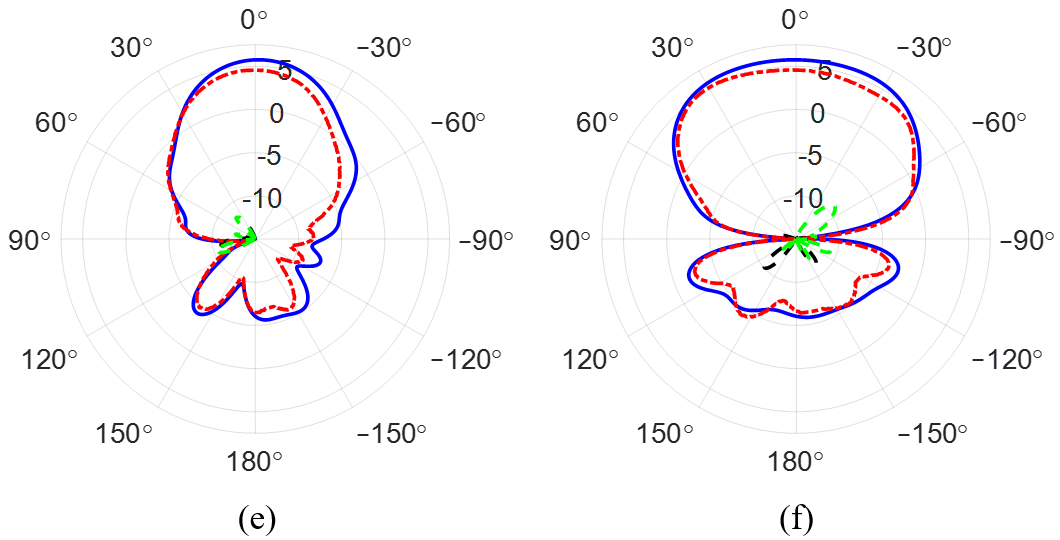}
	\vspace{0.5em}			
	\newline
	\includegraphics[width=3.1in]{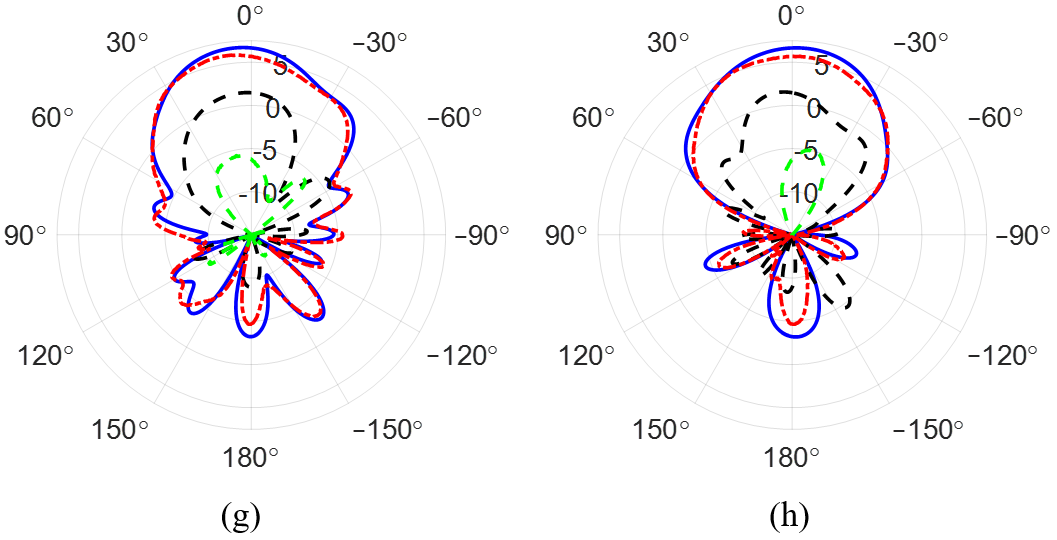}		
	\vspace{0.5em}	
	\newline
	\includegraphics[width=3.1in]{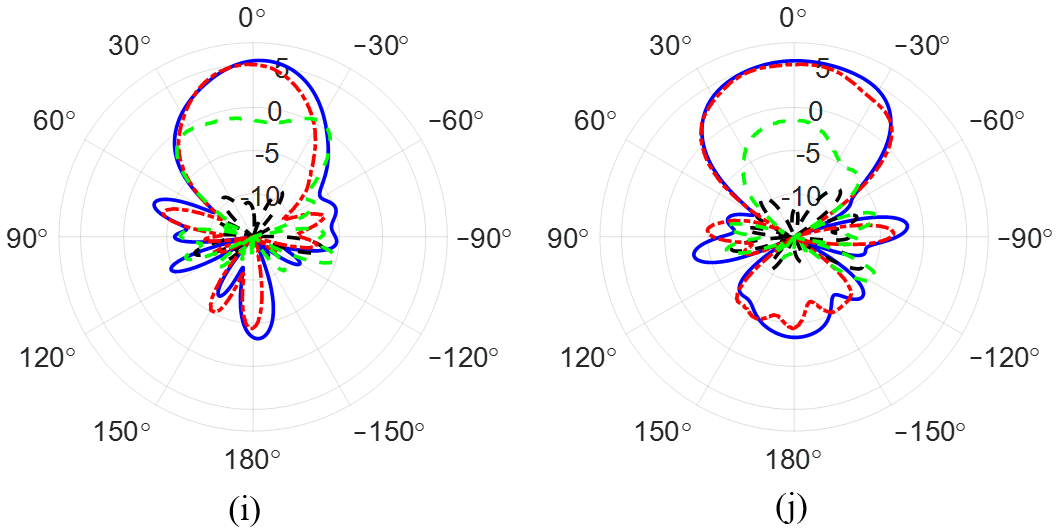}		
	\caption{Simulated and measured gain patterns of the UPR antenna: (a) E-plane at 1.2 GHz; (b) H-plane at 1.2 GHz; (c) E-plane at 1.6 GHz; (d) H-plane at 1.6 GHz; (e) E-plane at 2.1 GHz; (f) H-plane at 2.1 GHz; (g) E-plane at 2.6 GHz;  (h) H-plane at 2.6 GHz; (i) E-plane at 3.0 GHz; (j) H-plane at 3.0 GHz.}
	\label{fig102}
\end{figure}

\begin{figure}[!t]
	\centering		
	\includegraphics[width=2.6in]{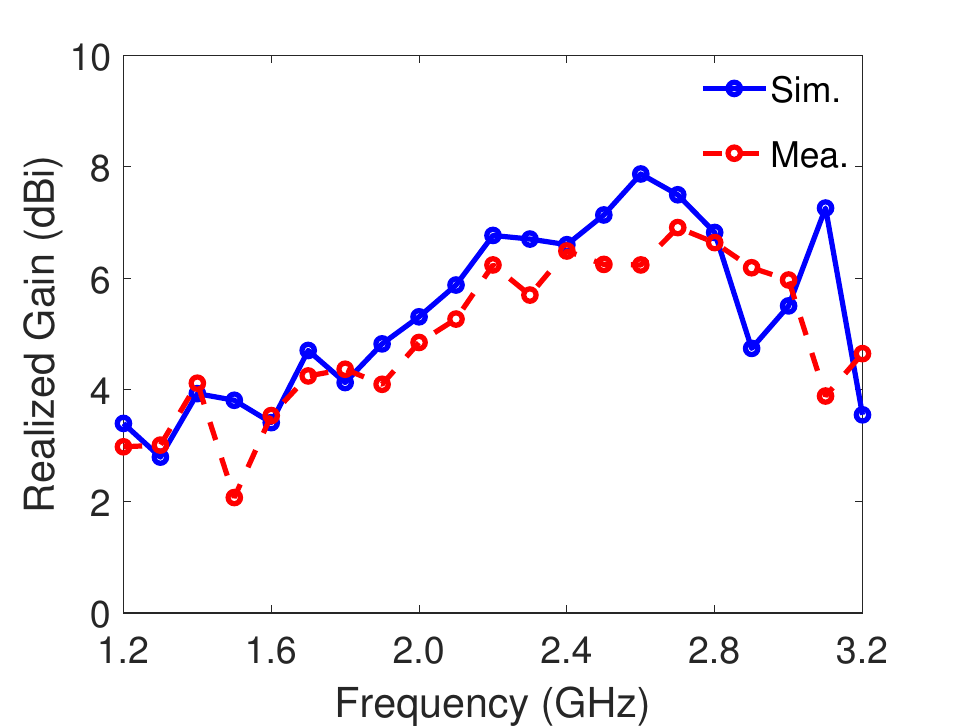}		
	\caption{Simulated and measured realized gain of the UPR antenna.}
	\label{fig103}
\end{figure}

The UPR antenna is simulated and optimized using Ansys HFSS. The individual components, including the Vivaldi radiators, power divider, rat-race coupler, and $90^\circ$ phase shifter, are first designed separately to operate over an ultrawide frequency range of 1.2–3.2~GHz. These components are then integrated and jointly optimized as a complete antenna structure, which is subsequently fabricated. This frequency range was selected as a representative frequency band for near-surface GPR sensing, including buried threats detection \cite{garciafernandez2023array}, landmine detection \cite{bestagini2021landmine}, shallow-buried pipe detection \cite{prego2017efficient}, and tree root detection and characterization \cite{liu2013tree}. The proposed antenna is intended to demonstrate the advantage of the frequency-dependent polarization-rotation characteristics for improving detection robustness across targets with different geometries and orientations. The antenna design can be scaled and optimized for other frequency ranges according to specific GPR application requirements.

The power divider and the pair of unequal-length 50-$\Omega$ microstrip lines, which generate the required frequency-dependent phase difference of $2\Psi$ for LP rotation, are first examined before presenting the overall antenna characteristics. Their combined simulated S-parameters are shown in Fig.~\ref{fig200}. In the figure, port 1 is the input and ports 2 and 3 are the outputs. The power divider achieves good impedance matching across the operating band, with $|S_{11}|$ below $-10$~dB, and provides nearly equal power division between the two output ports, as shown by $|S_{21}|$ and $|S_{31}|$ in Fig.~\ref{fig200}(a). Moreover, the phase difference between the two output ports, denoted as $2\Psi = \angle S_{21} - \angle S_{31}$, varies from $-180^\circ$ to $180^\circ$ over 1.2 – 3.2~GHz, as shown in Fig.~\ref{fig200}(b). According to (\ref{eq3}), the variation of $\Psi$ from $-90^\circ$ to $90^\circ$ causes the LP direction of the antenna to rotate continuously from $-90^\circ$ to $90^\circ$ as the frequency increases over the operating band, thereby achieving the desired UPR performance. For brevity, the detailed simulation results of other components are omitted, and only the final antenna performance is presented as follows.

The fabricated antenna prototype is shown in Fig.~\ref{fig100}, where the frequency-controlled feeding network is attached to the antenna structure using plastic screws. The $S$-parameters are measured using a Copper Mountain Technologies M5090 vector network analyzer (VNA) \cite{m5090_datasheet}. Fig.~\ref{fig101} presents the simulated and measured $|S_{11}|$ of the antenna. The simulated $|S_{11}|$ remains below $-7.5$~dB over most of the desired frequency range of 1.2–3.2~GHz, indicating that more than $82\%$ of the input power is delivered to the antenna and subsequently radiated. This corresponds to an ultrawide fractional bandwidth of $90.9\%$. At the worst matching point of 3.1~GHz, the measured $|S_{11}|$ is $-5.2$~dB, indicating that approximately $70\%$ of the input power is delivered to the antenna and subsequently radiated. This still shows effective radiation performance for GPR applications. The measured $|S_{11}|$ agrees well with the simulated result. The discrepancies are mainly attributed to fabrication tolerances and assembly inaccuracies. \par

The radiation patterns of the UPR antenna are measured in the Satimo StarLab near-field antenna measurement chamber at the University of Wisconsin–Madison. To characterize the polarization rotating performance, the LP inclination angle of the antenna is evaluated across the operating frequency band. Fig.~\ref{fig104} shows the simulated and measured LP inclination angles from 1.2 to 3.2~GHz. The simulated results show that the LP angle increases approximately linearly with frequency, achieving continuous polarization rotation from $2^\circ$ to $170^\circ$ over the operating band. The measured results show a rotation from $9.4^\circ$ to $171.28^\circ$. The slight discrepancies between the simulated and measured results could be attributed to fabrication and assembly tolerances, as well as small mounting misalignments during measurement. Despite these discrepancies, the fabricated antenna demonstrates the expected polarization rotating behavior.

The simulated and measured E- and H-plane radiation patterns at five representative frequencies: 1.2, 1.6, 2.1, 2.6, and 3.0 GHz are presented in Fig.~\ref{fig102}. 
The E-plane is defined as the plane formed by the LP direction and the z-axis, as shown in Fig.~\ref{fig11}. The H-plane is defined as the plane formed by the cross-polarization (XP) direction and the z-axis. Since the LP direction rotates with frequency, both the E- and H-planes are frequency dependent and thus rotating with frequency.
As shown in Fig.~\ref{fig102}, the measured radiation patterns agree reasonably with the simulated results. The measured E-plane half-power beamwidths (HPBWs) at the five frequencies are $164^\circ$, $61^\circ$, $60^\circ$, $72^\circ$, and $38^\circ$, respectively. The measured H-plane HPBWs are $165^\circ$, $93^\circ$, $104^\circ$, $73^\circ$, and $70^\circ$, respectively. At most frequencies, the measured XP level at boresight remains below $-$10~dB, except at 3.0~GHz, where an XP level of $-$6.8~dB is measured. Fig.~\ref{fig103} shows the simulated and measured peak gains. The measured realized gain varies from 2.1 to 6.4~dBi over the operating frequency band.

In summary, the developed UPR antenna demonstrates a near-linear rotation of the LP angle with increasing frequency over 1.2–3.2~GHz. This capability enables flexible LP states with a single-port excitation, making the antenna well suited for low-complexity GPR systems and improving the robustness of subsurface target detection. Its performance in detecting targets with different geometries is discussed in the following section.

\section{UPR Antenna-Enabled GPR System and Experimental Evaluation}
\label{sec:sectionIV}

To validate the capability of the developed UPR antenna to enhance GPR detection robustness, this section presents the UPR antenna-based GPR system and evaluates its detection performance for several representative subsurface scatterers.

\subsection{GPR System Design and Calibration}
\begin{figure}[t]
    \centering
    \begin{tabular}{c}
        \includegraphics[width=0.9\linewidth]{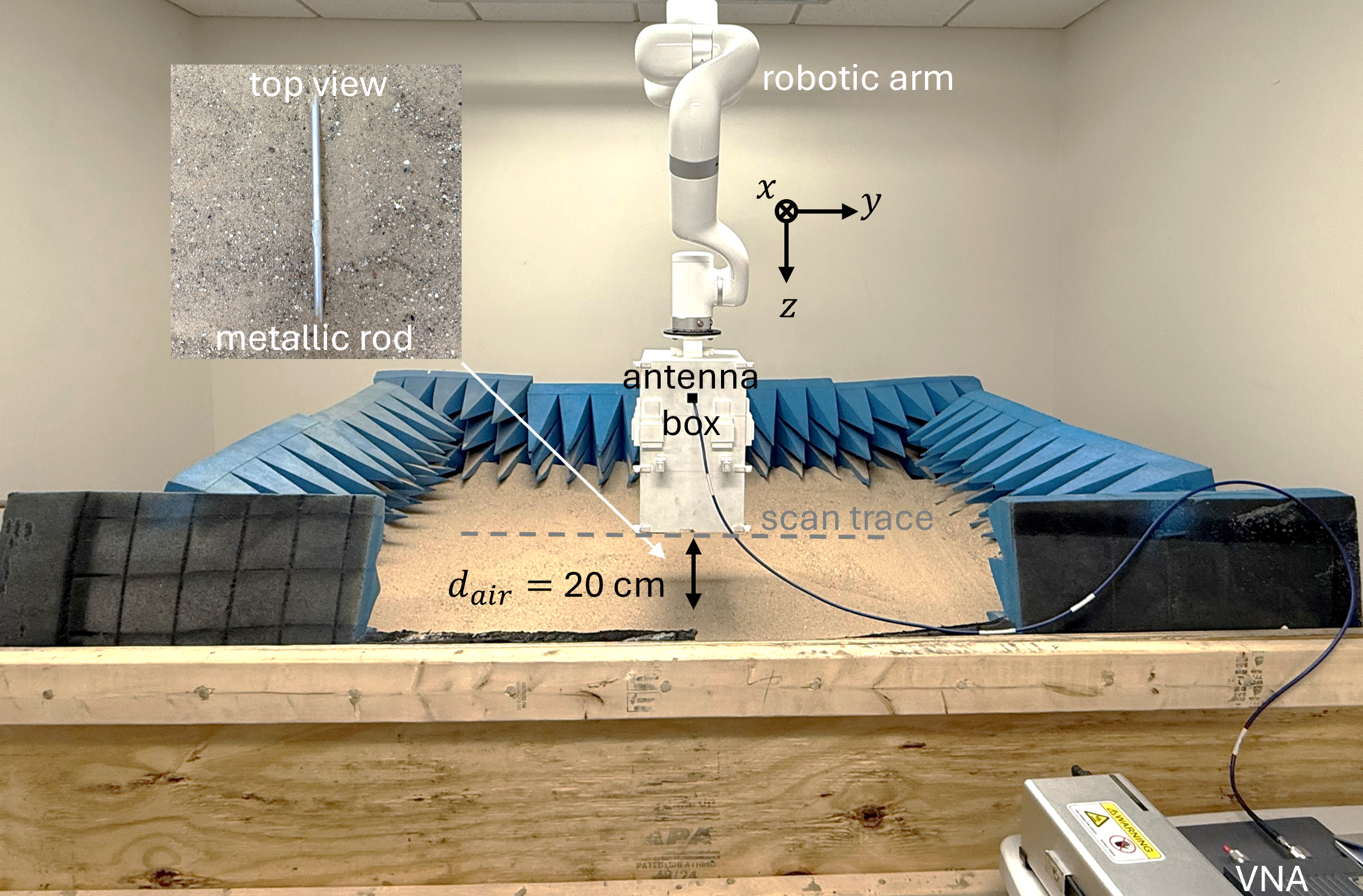} \\
        \footnotesize{(a)} \\
        
        \includegraphics[width=0.90\linewidth]{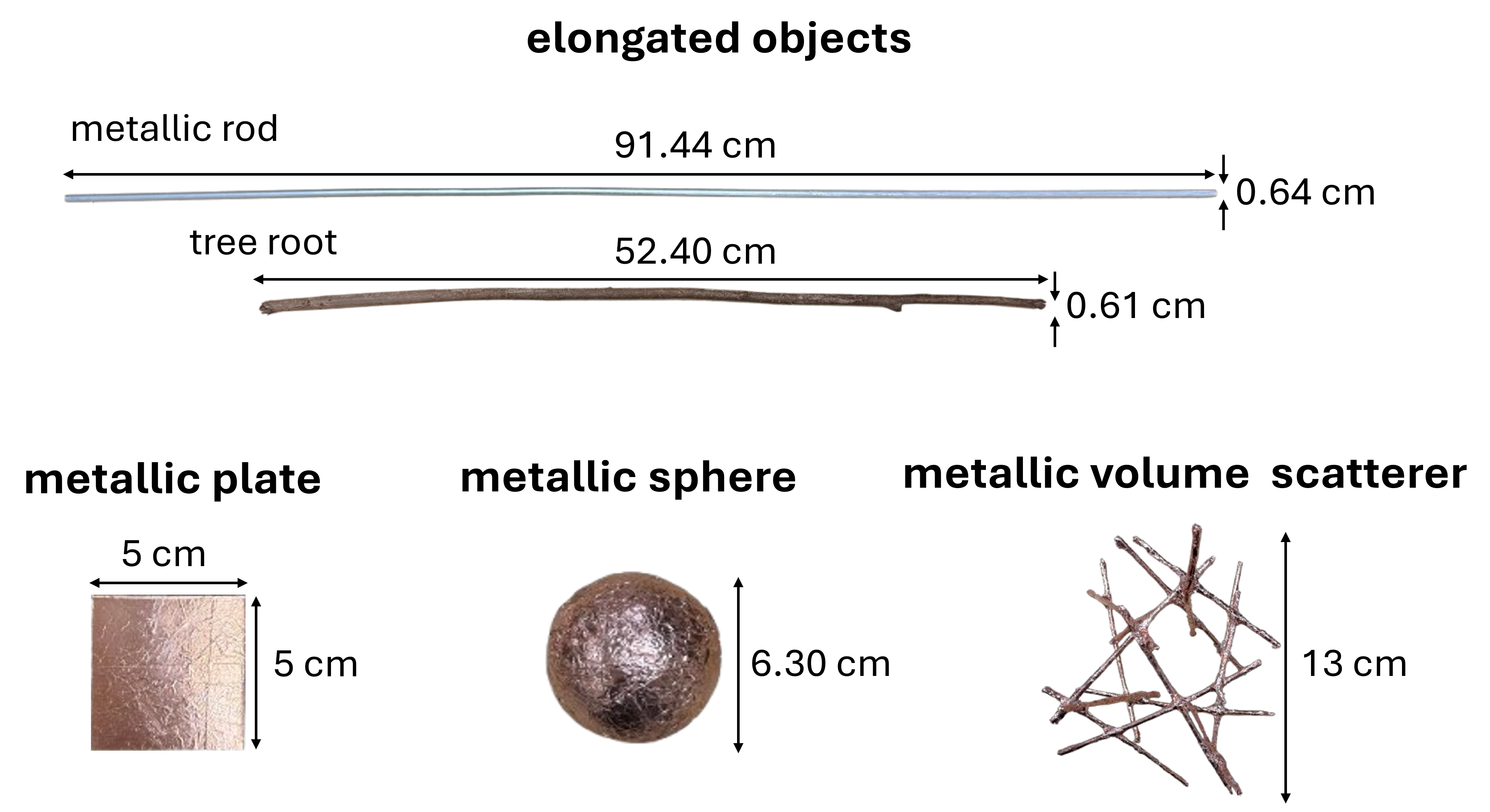} \\
        \footnotesize{(b)} \\

    \end{tabular}
    \caption{(a) Experimental setup for subsurface target detection using different GPR systems, where the antenna box is equipped with either the designed UPR antenna or the conventional SPLP antenna. (b) Objects of different geometries used in the experiment, including a metallic rod, tree root, metallic plate, metallic sphere, and metallic volumetric scatterer.}
    \label{fig:setup_and_object}
\end{figure}

A single-channel monostatic stepped-frequency continuous-wave GPR system is developed by connecting the designed UPR antenna to a VNA operating as a transceiver, as shown in Fig.~\ref{fig:setup_and_object}(a). The antenna is enclosed in an absorber-lined housing and mounted on a robotic arm to facilitate automated data acquisition. For comparison, a conventional single-channel GPR system employing a single-port fixed linearly polarized (SPLP) antenna (SPLP-based GPR) is also implemented under the same experimental conditions.

According to the Federal Communications Commission (FCC) definition, a system with an absolute bandwidth of at least 500 MHz or a fractional bandwidth of at least 20$\%$ qualifies as an ultrawideband system \cite{FCC_UWB}. In addition, IEEE Std 686-2024 defines an ultrawideband radar as having a fractional bandwidth greater than 25$\%$ \cite{IEEE686_2024}. The proposed GPR operates from 1.2 to 3.2 GHz, corresponding to a 2.0 GHz absolute bandwidth and a fractional bandwidth of 90.9$\%$, and therefore satisfies both ultrawideband criteria.

To ensure accurate target detection across the full operating band, a phase-delay calibration procedure is devised to compensate for the frequency-dependent phase response in the UPR antenna, and applied to all radar measurements. The antenna phase delay is obtained from a reference measurement, in which the antenna is placed at a known distance, $h$, from a large flat copper plate that is treated as a perfect electric conductor. The copper plate, with its conductivity of $5.8 \times 10^{7}\ \mathrm{S/m}$ and thickness of $0.635\ \mathrm{mm}$, closely approximates a perfect electric conductor (PEC), with an electric-field reflection coefficient of approximately $\Gamma = -1$. The plate therefore provides a well-defined and reproducible reflection suitable for antenna phase-delay calibration. The measured reflection phase from the copper plate, $\angle S_{\mathrm{PEC}}(f)$, includes the two-way antenna phase delay, $\phi_A(f)$, the round-trip propagation phase delay, $\phi_D(f) = 4\pi f h/c$, where $c$ is the speed of light, and the phase of the reflection coefficient of the copper plate, $\pi$. Therefore, the frequency-dependent two-way antenna phase delay can be obtained as: $\phi_A(f) = -\angle S_{\mathrm{PEC}}(f) - \phi_D(f) - \pi$.

The extracted antenna phase delay is subsequently removed from the measured frequency response of each A-scan to compensate for the intrinsic frequency-dependent phase distortion introduced by the antenna system.
After phase compensation, the frequency-domain data from 1.2 to 3.2 GHz are multiplied by a Kaiser window with $\beta=3$ to suppress spectral leakage and reduce sidelobe levels before applying an inverse Fourier transform to obtain the corresponding time-domain A-scans. The collection of calibrated A-scans is then assembled to form the final B-scan radargrams. The phase-delay calibration procedure is applied to both the developed UPR-based GPR and the SPLP-based GPR measurements to ensure a fair comparison. 

In GPR measurements, an A-scan is a one-dimensional radar trace acquired at a single antenna position along the scan path. A B-scan radargram is then formed by arranging a sequence of A-scans acquired at different antenna positions along the scan trace, where the horizontal axis represents the antenna position along the scan trace, the vertical axis represents time, and the color scale represents the signal magnitude.

\subsection{Experimental Setup}

The GPR measurements are performed by collecting the reflection coefficient $S_{11}$ from 1.2 to 3.2 GHz with 1001 points. The output power is set as 0 dBm, and the IF bandwidth is set to 1 kHz. The same system setup is used for the GPR measurements with both the UPR antenna and the SPLP antenna. As shown in Fig. \ref{fig:setup_and_object}(a), the UPR antenna-based GPR and the SPLP-based GPR are mounted on a robotic arm to perform B-scan measurements over a distance of 80 cm with a step size of 1 cm, with the antenna-to-sand-surface distance $d_{\mathrm{air}}$ = 20 cm, for detecting buried targets in a sandy environment. The 1-cm B-scan step size is selected to satisfy the spatial Nyquist sampling criterion and avoid spatial aliasing. The criterion requires a spatial sampling interval of $\Delta x \leq \lambda_{\min}/2$, where $\lambda_{\min}$ is the minimum wavelength in the medium. At the highest operating frequency of 3.2 GHz and with the relative permittivity of the sand of approximately $\varepsilon_r=3$, the maximum allowable step size is approximately 2.70 cm. Therefore, the selected step size satisfies the criterion and ensures fine spatial sampling. Both antennas are placed in a box lined with absorbers to reduce environmental noise. The target burial depth is $d$. \par

Fig.~\ref{fig:setup_and_object}(b) shows the five objects of different geometries used in the experiment, which include two elongated objects (a metallic bar and a tree root that is soaked in water overnight), a flat metallic plate, a sphere, and a volume scatterer with multiple branches. They represent commonly encountered targets in different GPR applications \cite{feng_subsurface_2015}.

\begin{figure}[!t]
	\centering
	\begin{tabular}{c}
		\includegraphics[width=0.95\linewidth]{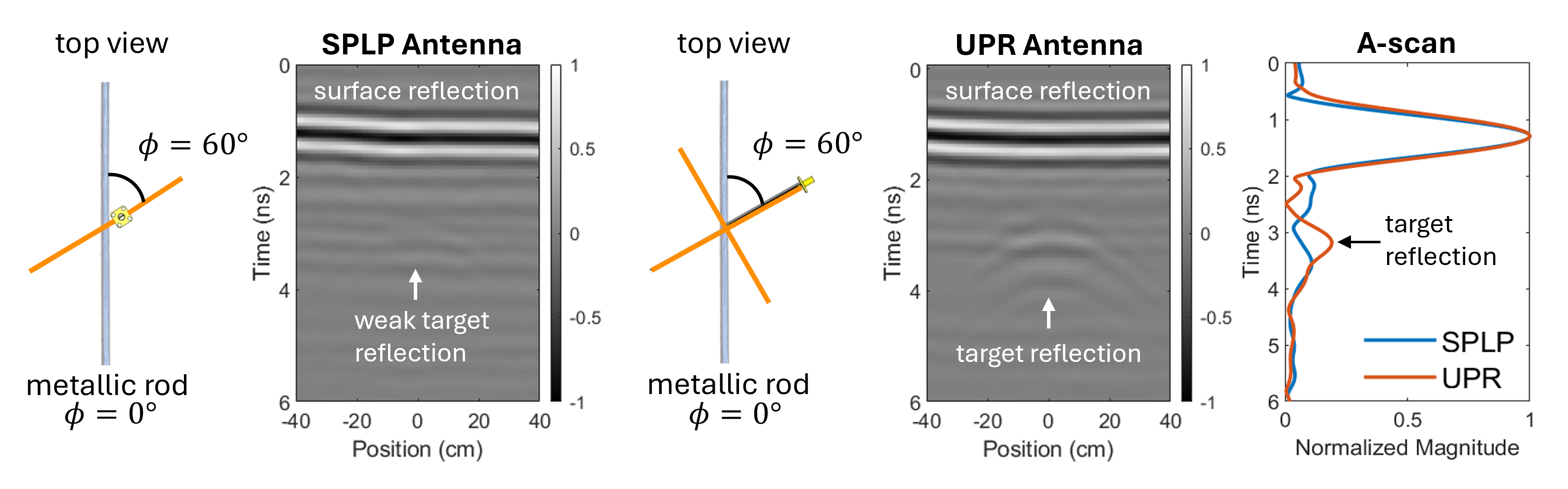} \\
		\footnotesize{(a)} \\
		
		\includegraphics[width=0.95\linewidth]{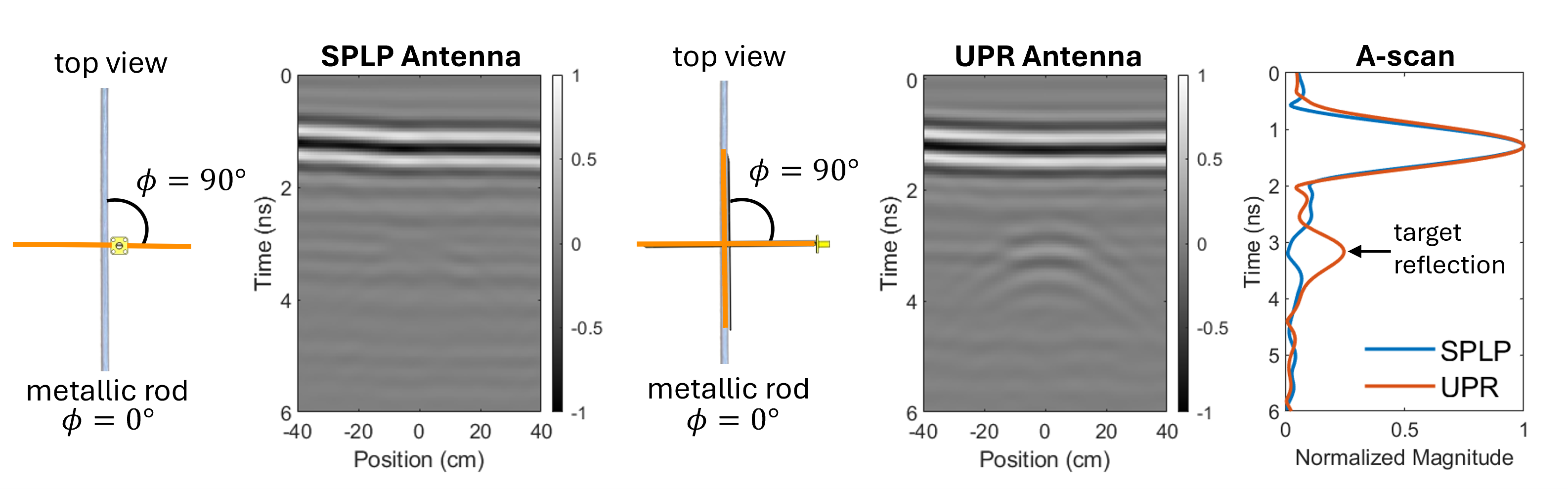} \\
		\footnotesize{(b)} \\
		
		\includegraphics[width=0.95\linewidth]{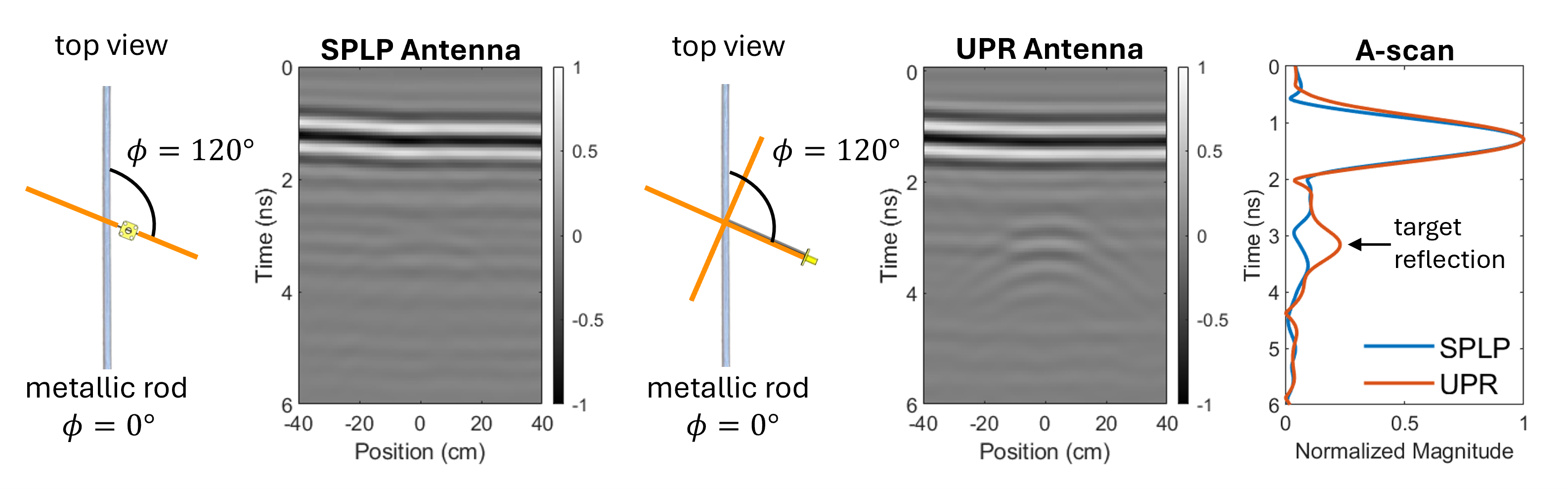} \\
		\footnotesize{(c)} \\
		
	\end{tabular}
	\caption{Comparisons of B-scans acquired by the SPLP- and UPR-based GPR systems for a metallic rod placed along the $x$-axis and buried at a depth of $d = 17.2$~cm, along with A-scans comparison of the center trace when the antenna is right above the metallic rod. The antenna-to-sand-surface distance is $d_{\mathrm{air}} = 20$~cm, and the antenna port axis is oriented at (a) $\phi = 60^\circ$, (b) $\phi = 90^\circ$, and (c) $\phi = 120^\circ$.}
	\label{fig:rod}
\end{figure}

Elongated objects depolarize electromagnetic waves based on their orientation, making reliable detection of randomly oriented elongated objects difficult for conventional single-polarized radar systems, especially when the antenna polarization is nearly perpendicular to the object axis \cite{daniels2004ground, liu2019hybrid}. Therefore, for the two elongated objects in this study, the objective is to compare the detection performance of the UPR- and SPLP-based GPR systems at different object orientations. In the experiment, the elongated objects were buried along the $x$-axis where $\phi = 0^\circ$. To measure the target responses at different relative orientations, the targets were maintained at fixed orientations, and the antenna box was physically rotated using the robotic arm. This physical rotation was used only to control the relative angular orientation between the antenna and the target for experimental convenience and is independent of the frequency-dependent polarization rotation inherently produced by the UPR antenna. Equivalently, the same relative orientations could be obtained by keeping the antenna stationary and rotating the target. Specifically, the antenna port axes of both the UPR and SPLP antennas were oriented at $\phi=60^\circ$, $\phi=90^\circ$, and $\phi=120^\circ$, respectively, as shown in Figs.~\ref{fig:rod} and ~\ref{fig:root}. For other objects with weak orientation-dependent scattering characteristics, the antenna port axes were all oriented at $\phi=0^\circ$, as shown in Fig.~\ref{fig:other_objects}. During each measurement, B-scan data are collected by moving the antenna box along an 80-cm scanning path, with the target placed at the center of the scan. 
\par

\subsection{Experimental Results for Different Buried Targets}
\label{sec:experimental result}

\subsubsection{Comparison for Elongated Objects}
For the elongated metallic rod with a diameter of \(D = 0.64~\mathrm{cm}\) and a length \(L = 91.44~\mathrm{cm}\), the transverse electrical size, \(ka = \frac{\pi D}{\lambda}\), increases from 0.14 to 0.37 over the antenna's 1.2--3.2~GHz operating band. Therefore, the transverse scattering of the rod is in the low-frequency region (\(ka \ll 1\)) \cite[pp. 209--213]{ruck_rcs_vol1}. Its longitudinal electrical dimension, \(L/\lambda\), increases from 6.35 to 16.93, confirming that the metallic rod is electrically long throughout the operating band. The scattering response of such an elongated target is strongly polarization dependent, with a stronger response when the incident electric field is aligned with the target axis and a reduced response as the LP becomes increasingly misaligned \cite[pp. 209--213]{ruck_rcs_vol1}. This analysis also applies qualitatively to the tree root, which is similarly electrically small in diameter (0.61~cm) and electrically long along its length (52.40~cm). \par
As shown in Fig.~\ref{fig:rod}, the B-scans and the center-trace A-scans are compared for the metallic rod, a common GPR target in civil engineering applications, placed along the $x$-axis ($\phi = 0^\circ$) and buried at a depth of $d= 17.20$ cm. The SPLP- and UPR-based GPRs are tested with the antenna port axis orientated at  $\phi = 60^\circ$, $90^\circ$, and $120^\circ$, respectively. Each B-scan and A-scan contains both the sand-surface reflection and the target reflection, and is normalized for comparison. 

The results show a clear difference between the two systems. For the SPLP-based GPR, the target response is very weak and difficult to distinguish in the demonstrated cases where the polarization direction is not aligned with the axis of the metallic rod, as can be seen from both the B-scan and center-trace A-scan comparisons in Fig.~\ref{fig:rod}. This behavior is expected for elongated targets because their scattering response strongly depends on the relative orientation between the target axis and the incident electric field. In contrast, the UPR-based GPR provides clear target reflections for all three orientations. This improvement is attributed to the frequency-dependent polarization rotation of the UPR antenna, which allows the radar to capture the backscattering response of elongated targets over a range of LP directions within a single measurement, with some polarization directions closely aligned with the bar axis. Therefore, the UPR antenna provides more stable detection performance than the SPLP antenna for elongated targets with different orientations. \par

The target burial depth can be subsequently estimated using the A-scan obtained when the antenna is directly above the metallic rod. In dry sand with relative permittivity of $\varepsilon_r=3$, the target depth is calculated by $d = \frac{c\Delta t}{2\sqrt{\varepsilon_r}}$, where $\Delta t$ is the two-way travel time. For the UPR-based GPR system with antenna port axis at $\phi = 60^\circ$, $90^\circ$, and $120^\circ$, the estimated depths are 16.91 cm, 17.18 cm, and 16.65 cm, respectively. These values agree well with the actual burial depth of 17.20 cm. The small depth error can be caused by small uncertainties in the rod placement and local heterogeneity in the sand.  \par

\begin{figure}[!t]
	\centering
	\begin{tabular}{c}
		\includegraphics[width=0.95\linewidth]{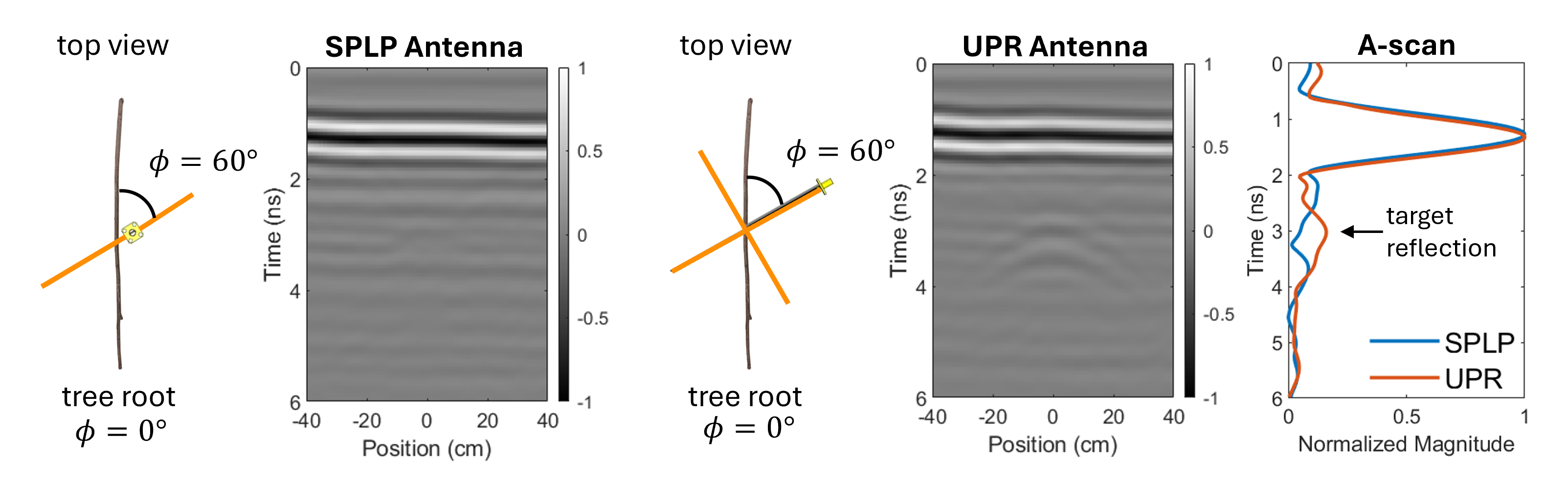} \\
		\footnotesize{(a)} \\
		
		\includegraphics[width=0.95\linewidth]{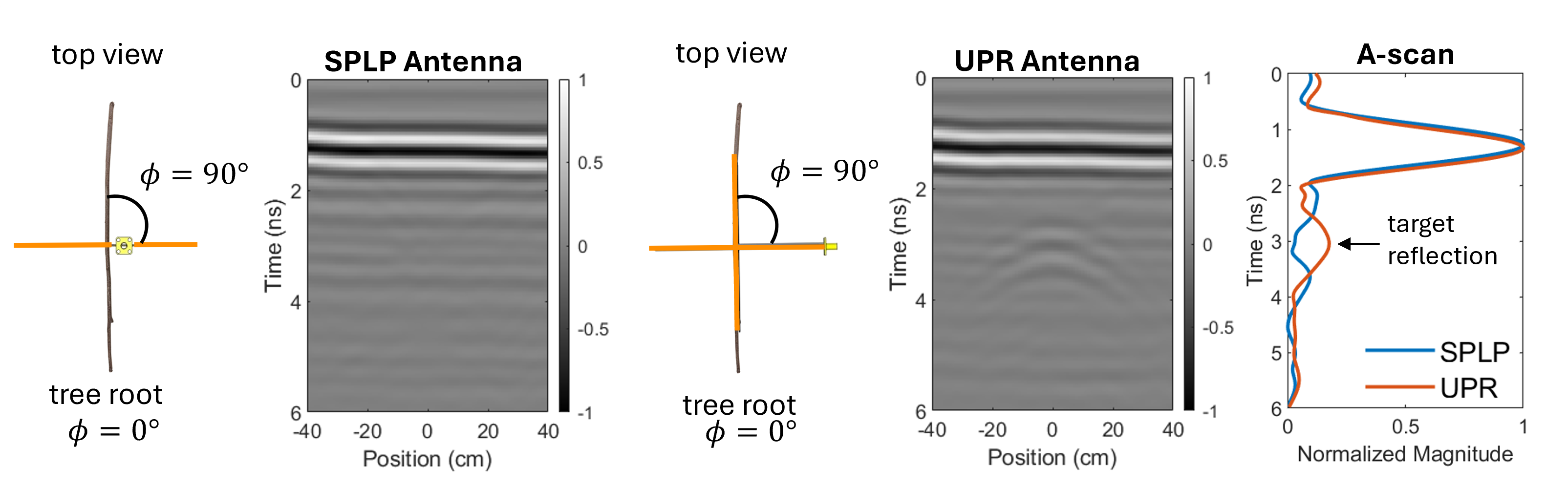} \\
		\footnotesize{(b)} \\
		
		\includegraphics[width=0.95\linewidth]{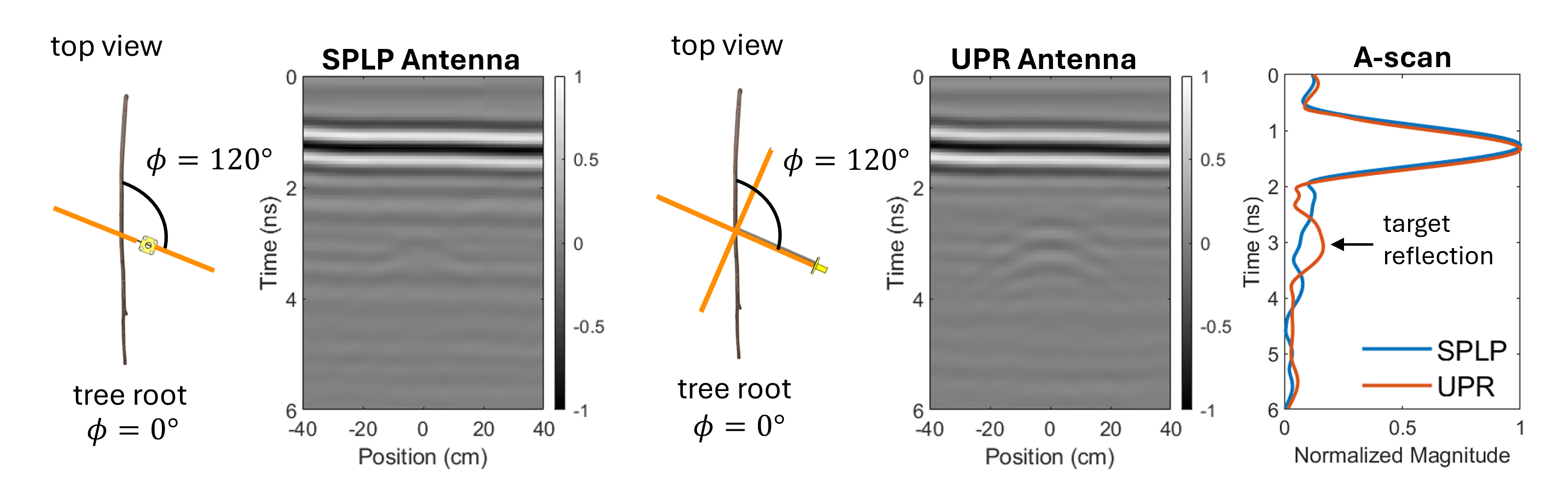} \\
		\footnotesize{(c)} \\
		
	\end{tabular}
	\caption{Comparisons of B-scans acquired by the SPLP- and UPR-based GPR systems for a tree root placed along the $x$-axis and buried at a depth of $d = 16.2$~cm, along with A-scans comparison of the center trace when the antenna is right above the tree root. The antenna-to-sand-surface distance is $d_{\mathrm{air}} = 20$~cm, and the antenna port axis is oriented at (a) $\phi = 60^\circ$, (b) $\phi = 90^\circ$, and (c) $\phi = 120^\circ$.}
	\label{fig:root}
\end{figure}

The same comparison is also performed for the tree root, which represents a common elongated subsurface target in environmental engineering. The tree root branch is placed along the  $x$-axis ($\phi = 0^\circ$) and buried at a depth of $d= 16.2$~cm. Both B-scan measurements are again obtained for UPR- and SPLP-based GPRs with antenna port-axis orientations of $\phi = 60^\circ$, $90^\circ$, and $120^\circ$. The experimental setup, measured B-scan, and center-trace A-scans are shown in Fig.~\ref{fig:root}. Similar to the metallic rod case, the SPLP-based GPR yields weak and poorly distinguishable target hyperbolas when the antenna polarization is misaligned with the root axis. In particular, as shown in Fig.~\ref{fig:root}(b), the root reflection becomes too weak to be reliably identified, resulting in unsuccessful target identification. In contrast, the UPR-based system consistently produces stronger root responses, resulting in clearer root reflections in the B-scans for all three tested orientations. \par

Overall, the results with elongated targets confirm that the UPR antenna improves the robustness of GPR detection for elongated targets. Compared with the SPLP antenna, the UPR antenna is less sensitive to target orientation because of its unique polarization variation across the frequency band. This characteristic is particularly favorable when the orientation of buried elongated objects is unknown, as is often the case in practical GPR surveys.

\begin{figure}[!t]
	\centering
	\begin{tabular}{c}
		\includegraphics[width=0.95\linewidth]{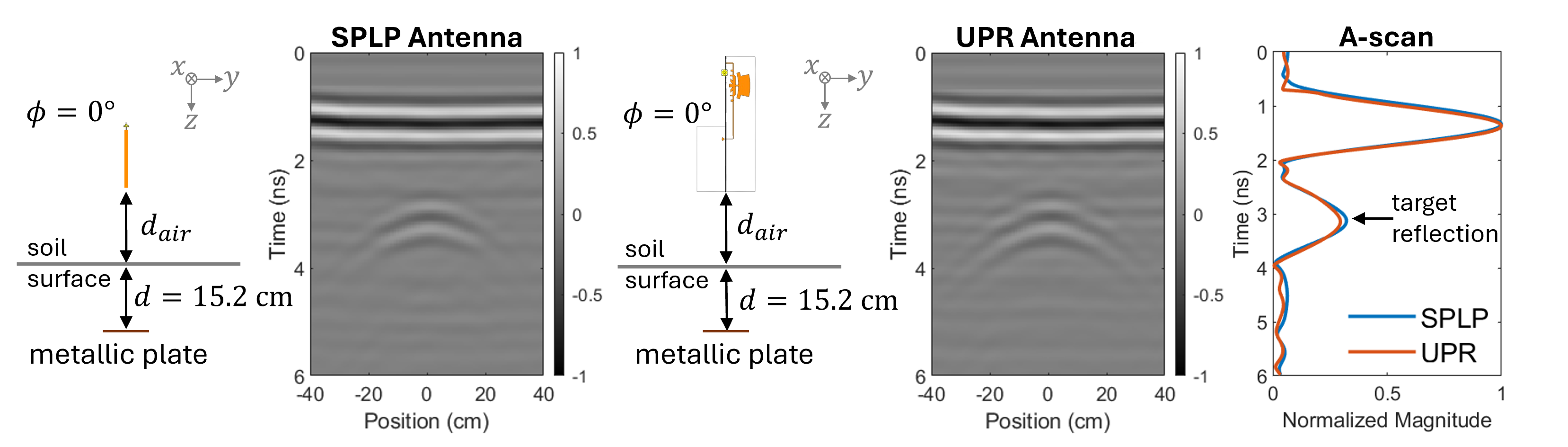} \\
		\footnotesize{(a)} \\
		
		\includegraphics[width=0.95\linewidth]{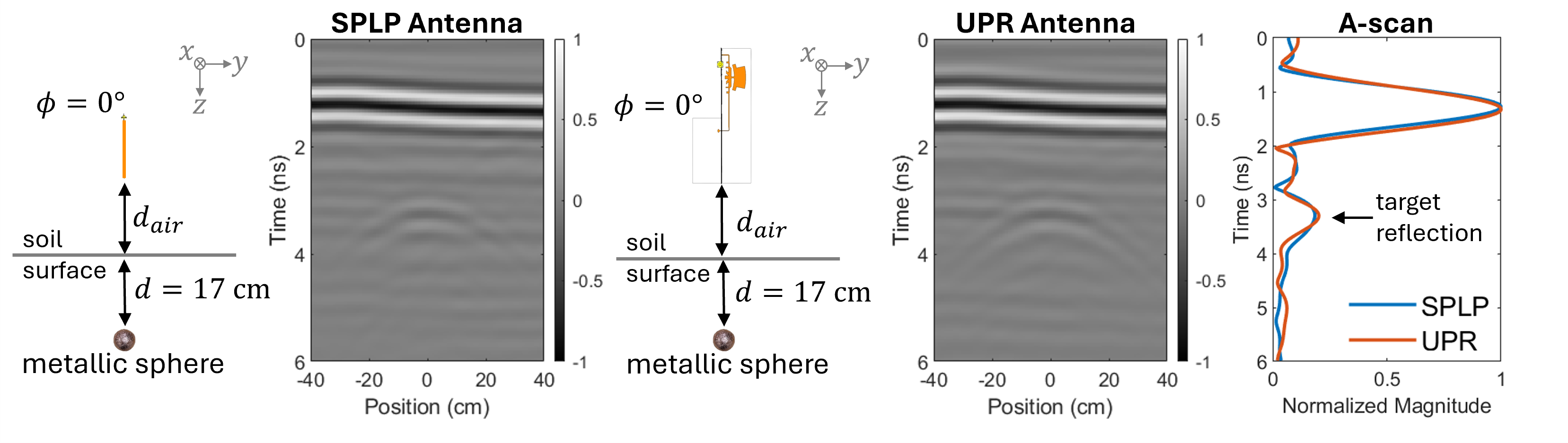} \\
		\footnotesize{(b)} \\
		
	\end{tabular}
	\caption{
		Comparisons of B-scans and A-scans of center trace acquired by the SPLP- and UPR-based GPR systems for (a) a plate and (b) a sphere.}
	\label{fig:other_objects}
\end{figure}

\subsubsection{Comparison for Other Metallic Targets}
The detection performance of two GPR systems is also compared for targets with less pronounced orientation-dependent scattering, including a metallic plate and a metallic sphere. The experimental configuration and the corresponding B-scan and center-trace A-scan results are presented in Fig.~\ref{fig:other_objects}.
The metallic plate has lateral dimensions of \(5\times5\) cm, with its lateral electrical size \(L/\lambda\) increasing from 0.35 to 0.93 over the operating band. Therefore, the scattering of the plate transitions from low-frequency to resonance region, and its intrinsic scattering response is frequency dependent. Under normal incidence, the planar conducting surface produces a predominantly specular reflection that is insensitive to the LP angle of the incident wave \cite{ruck_rcs_vol2}. For the metallic sphere with a diameter of 6.30 cm, the electrical size $ka$ increases from 1.37 to 3.67 over the operating band. This range places the sphere within the resonance region ($0.4 \leq ka \leq 20$) \cite[pp. 145--153]{ruck_rcs_vol1}, where its intrinsic backscattering response varies with frequency. Given that the sphere is rotationally symmetric, its backscattering response is insensitive to the LP angle of the incident wave. \par

\begin{figure}[t]
	\centering
	\begin{tabular}{c}
		\includegraphics[width=0.90\linewidth]{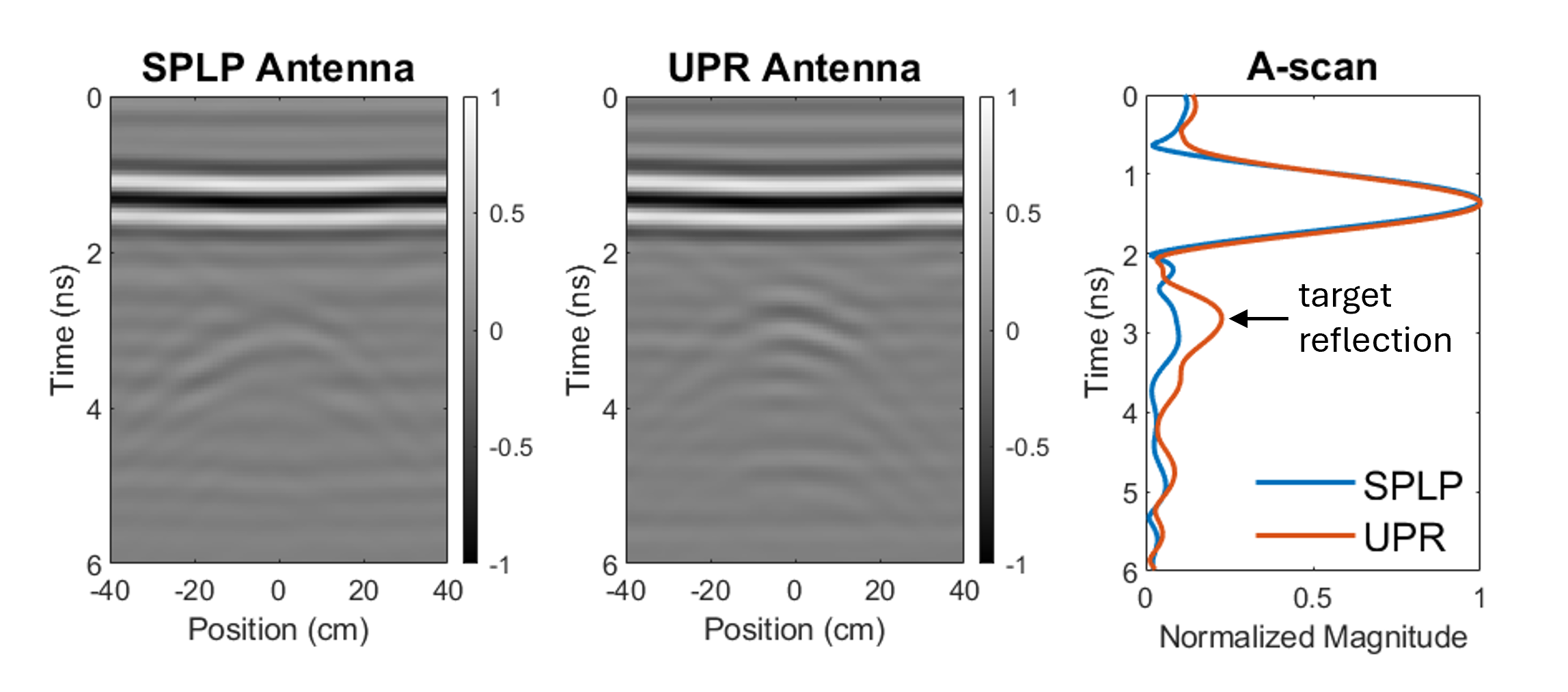} \\
		\footnotesize{(a)} \\
		
		\includegraphics[width=0.90\linewidth]{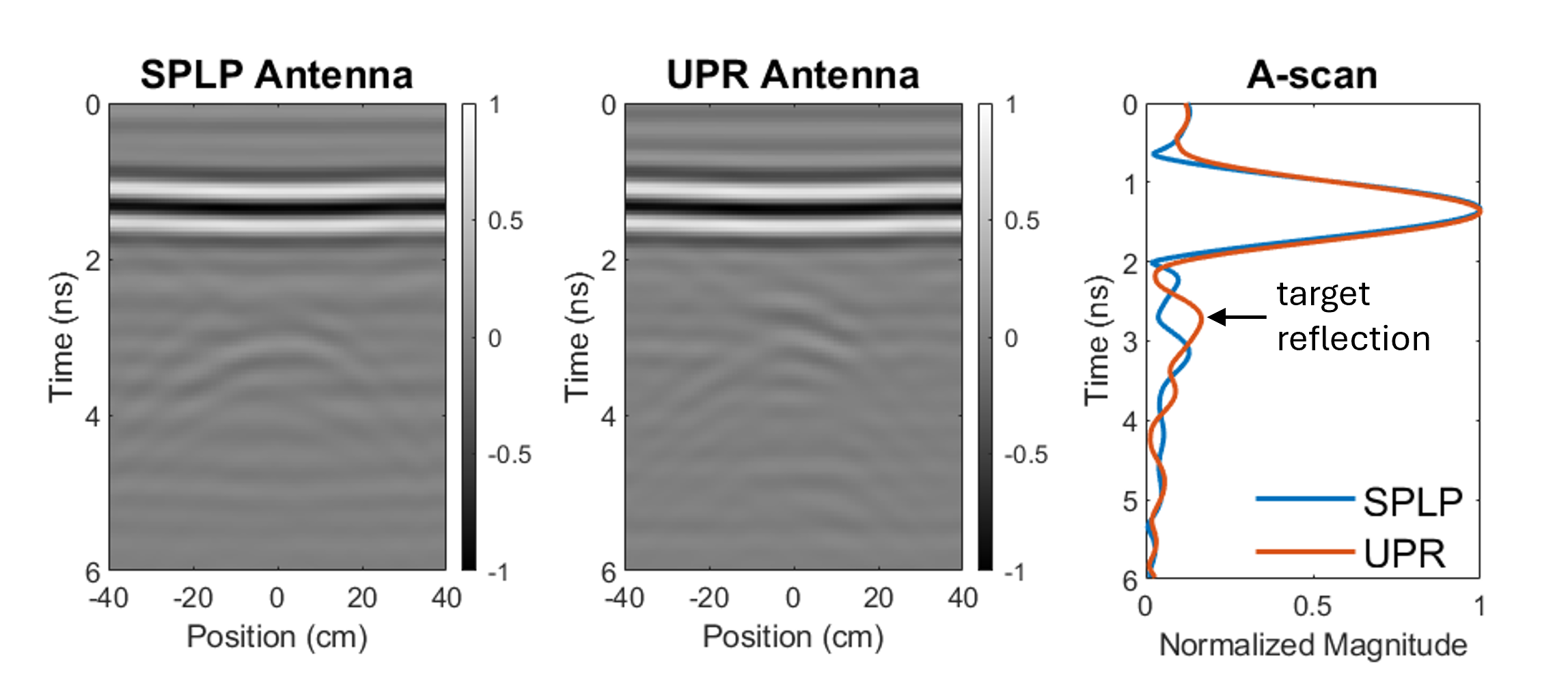} \\
		\footnotesize{(b)} \\
		
		\includegraphics[width=0.90\linewidth]{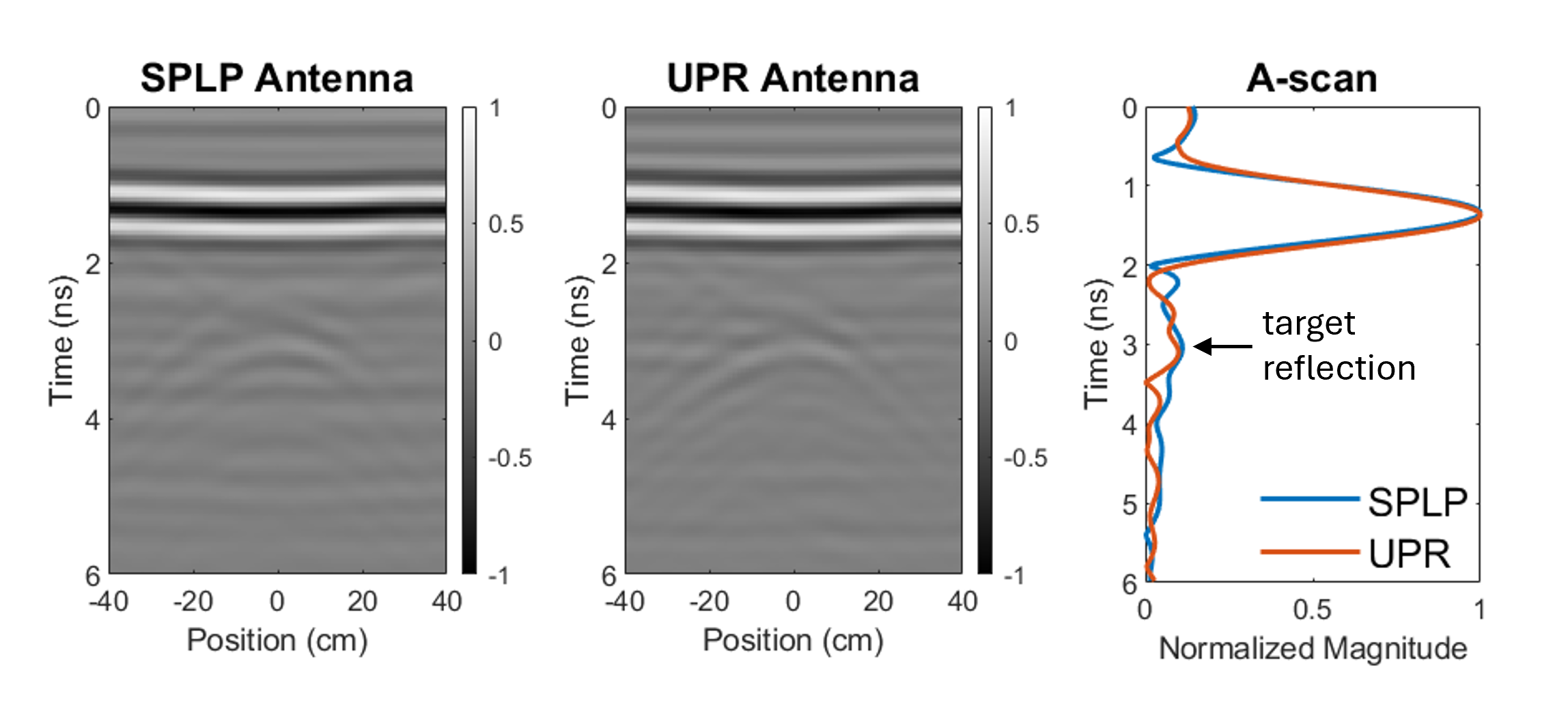} \\
		\footnotesize{(c)} \\

        \includegraphics[width=0.90\linewidth]{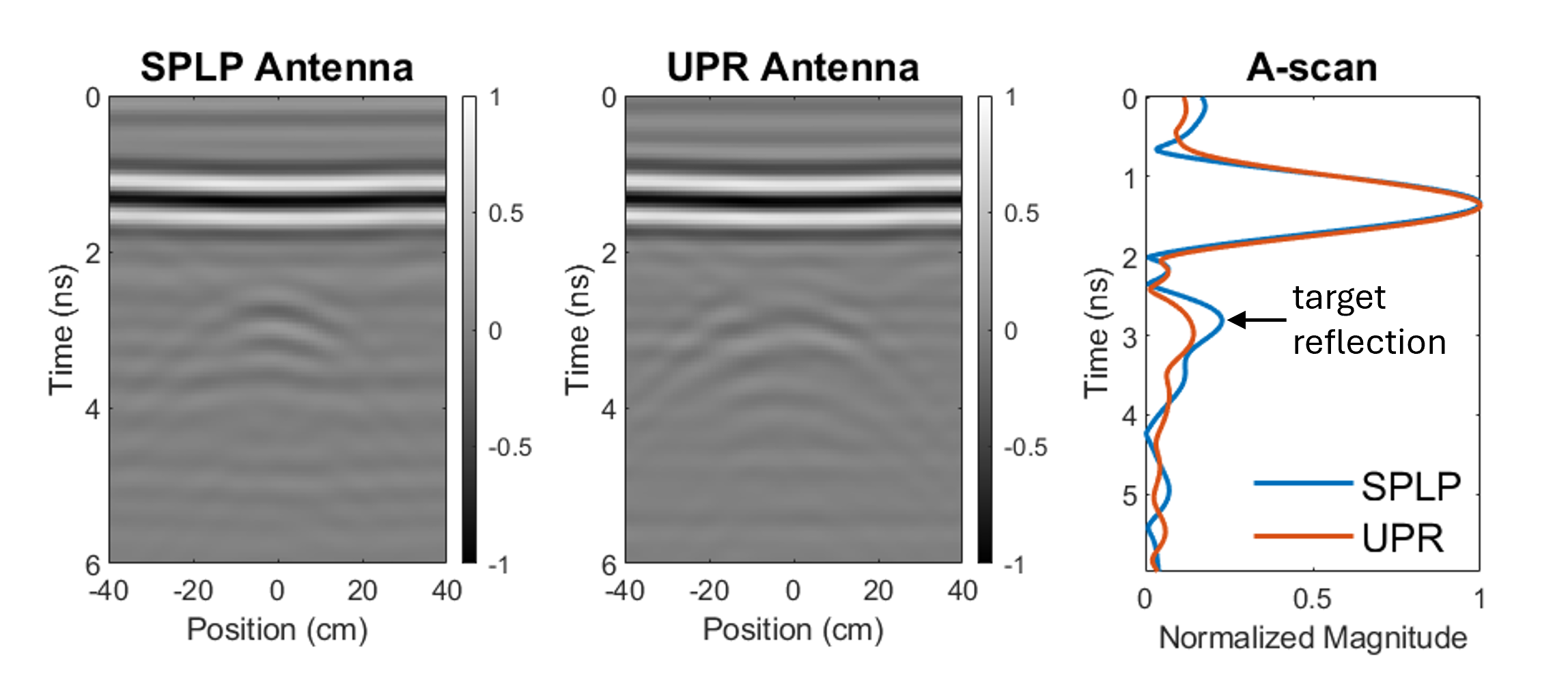} \\
		\footnotesize{(d)} \\
		
	\end{tabular}
	\caption{
		Comparisons of B-scans acquired by the SPLP- and UPR- based GPR systems for a metallic volume scatterer buried at a depth of $d = 13$~cm, along with A-scans comparison of the center trace when the antenna is right above the metallic volume scatterer. The antenna-to-sand-surface distance is $d_{\mathrm{air}} = 20$~cm, and the antenna port axis is oriented at (a) $\phi = 0^\circ$, (b) $\phi = 60^\circ$, (c) $\phi = 90^\circ$, and (d) $\phi = 120^\circ$.}
	\label{fig:scatterer}
\end{figure}

In addition to these cases, the metallic volume scatterer presented in Fig.~\ref{fig:setup_and_object}(b) was also experimentally investigated at four orientations relative to the antenna port axis: $\phi = 0^\circ$, $60^\circ$, $90^\circ$, and $120^\circ$, with the same burial depth of $d= 13$~cm and antenna-to-sand-surface distance of $d_{\mathrm{air}} = 20$~cm. The B-scan and center-trace A-scan results are presented in Fig.~\ref{fig:scatterer}. At $\phi = 0^\circ$ and $60^\circ$, the UPR antenna produces stronger target responses; at $\phi = 90^\circ$, the two antennas produce comparable responses; and at $\phi = 120^\circ$, the SPLP antenna produces a stronger response, but the target remains visibly detectable with the UPR antenna. These results show that the relative responses of the two systems vary with the placement and orientation of the volumetric scatterer and, overall, the UPR antenna provides more consistent detection across the tested orientations. \par
Overall, the UPR-based GPR performs well across all tested target geometries. Its main advantage is most evident for targets with orientation-dependent scattering, where polarization mismatch can significantly reduce the detection capability of conventional single-polarized GPR systems. 
By providing frequency-dependent LPs through a single-port antenna configuration, the UPR antenna enables more robust subsurface target detection without increasing the hardware complexity of conventional single-channel GPR systems. \par

The primary objective of these controlled experiments was to compare the detection capabilities of the UPR- and SPLP-based GPR systems under identical target and measurement conditions. The known target locations and burial depths were used only as ground truth for comparison with the measured results. In practical GPR surveys where the target location is unknown a priori, target identification relies on the spatial and temporal characteristics of the measured response. For a localized buried target, measurements acquired at successive antenna positions along a scan trace typically form a hyperbolic signature in the B-scan, as observed in Figs. ~\ref{fig:rod}--\ref{fig:scatterer}. In contrast, soil heterogeneity and other environmental factors may introduce additional clutter with different spatial characteristics, such as laterally extended responses. Conventional GPR processing and detection methods, such as hyperbolic-signature detection \cite{dou2017hyperbola},\cite{gamba2000neural}, can therefore be applied to identify characteristic target responses and facilitate target localization.\par

Note that the target detection results presented in this section were each obtained from a single B-scan measurement. To quantitatively assess the reliability of the measurements, a detailed uncertainty analysis based on a representative experimental configuration will be performed in Section V-B.
 
\section{Discussions}
In this section, we present a comparison with state-of-the-art GPR systems, demonstrate GUM Compliant Type A/Type B analysis, provide theoretical discussion for the comparison of the proposed UPR antenna with a single-port circular polarization (CP) antenna, and discuss the effects of effective bandwidth for orientation dependent targets.

\subsection{Comparison With State-of-the-Art GPR Systems}

To benchmark the developed system against state-of-the-art designs, Table~\ref{com_tab} provides a structured comparison in terms of operating frequency range, antenna type, polarization capability, number of RF channels, number of acquisitions required per detection, detection robustness, and system complexity.

As shown in Table~\ref{com_tab}, conventional polarization-diverse GPR systems generally achieve robust detection through either multi-channel hardware or repeated measurements with antenna reorientation. The full-polarimetric systems in \cite{luo2024automatic} and \cite{zhou2022multiparameter} acquire the HH, HV, VH, and VV scattering components, providing comprehensive polarimetric information at the expense of increased hardware and channel complexity. The dual-polarized systems in \cite{sun2023diameter} and \cite{liu2023improved} acquire complementary responses through two orthogonal polarization channels, using a dual-port Vivaldi antenna and a commercial dual-polarized GPR system, respectively. These configurations improve target characterization but require dedicated dual-polarized hardware and multiple RF channels. Alternatively, \cite{sun2021dual} and \cite{dong2023super} rely on repeated scans with antenna reorientation. Specifically, \cite{sun2021dual} performs dual-cross-polarized measurements at two antenna orientations separated by 45°, whereas \cite{dong2023super} uses a single-polarized GPR antenna at 0°, 45°, and 90° orientations to reconstruct full-polarimetric data. These repeated measurements increase the acquisition effort and may introduce inconsistencies among polarization datasets.

In contrast, our developed UPR-based GPR provides continuous LP rotation across $0^{\circ}$--$170^{\circ}$ within a single wideband acquisition using a single RF channel, without any mechanical reconfiguration. This enables consistent and polarization-diverse detection of orientation-sensitive subsurface targets while maintaining low hardware complexity, representing a practical advancement over conventional polarimetric GPR systems.

\begin{table*}[!t]
	\renewcommand{\arraystretch}{1.5}   
	\centering
	\footnotesize
	\caption{Comparison of the Proposed UPR Antenna-Enabled GPR System with State-of-the-Art Designs}
    \label{com_tab}
    \begin{threeparttable}
	\begin{tabular}{c|c|c|c|c|c|c|c|c}
		\hline
		\hline
		Ref.  & \makecell[c]{ Frequency \\ range (GHz)} & \makecell[c]{ Antenna \\ type} & \makecell[c]{GPR polarization \\ capability} & \makecell[c]{No. of  Tx\\ channels} & \makecell[c]{No. of  Rx\\ channels} & \makecell[c]{No. of \\ measurements$^\dagger$} & \makecell[c]{Detection \\robustness} & \makecell[c]{Hardware\\ complexity} \\
		\hline
		  \cite{luo2024automatic}  & 0.5 -- 6.5 & \makecell[c]{Dual-polarized \\ Vivaldi} & Full polarimetric  & 2 & 2 & 1 &  High &  High \\
        \cite{sun2021dual}  & 0.4 -- 4.0 & \makecell[c]{Dual-polarized \\ Vivaldi} & \makecell[c]{Dual \\ cross-polarization}  & 1 & 1  & 2  &  High &  Medium \\
        \cite{sun2023diameter}  & 0.5 -- 3.3 & \makecell[c]{Dual-polarized \\ Vivaldi} & Dual polarization  & 2 & 2 & 1 &  Medium  &  Medium\\
        \cite{zhou2022multiparameter}  & 0.8 -- 4.0 & \makecell[c]{Dual-polarized \\ Vivaldi} & Full polarimetric  & 2  & 2 & 1 &  High &  High\\
        \cite{dong2023super}  & 1.0 -- 2.0 & N.A. & Single polarization  & 1 & 1  & 3 &  High &  Low \\
        \cite{liu2023improved}  & 2.0$^\S$ & N.A. & Dual polarization  & 2 & 2 & 1 &  Medium &  Medium \\
        This work & 1.2 -- 3.2 & \makecell[c]{Frequency-\\dependent Vivaldi} & \makecell[c]{Single-port continuous \\ LP rotation} & \makecell[c]{1$^{*}$} & \makecell[c]{1$^{*}$} & 1 & High & Low \\
		\hline    
		\hline
	\end{tabular}
    \begin{tablenotes} \footnotesize \item N.A.: Not available; Tx: Transmit; Rx: Receive; $^{*}$: shared Tx/Rx channel; $^\S$: Center frequency, bandwidth not reported; $^\dagger$: Number of measurements required to obtain the reported detection capability or polarization information.
    \end{tablenotes}
    \end{threeparttable}
\end{table*}

\subsection{Uncertainty Analysis}
To investigate the measurement uncertainty of the proposed GPR system, an uncertainty analysis was performed following the Guide to the Expression of Uncertainty in Measurement (GUM) framework \cite{JCGM2008}. The analysis considered both Type A and Type B uncertainty evaluations.

\subsubsection{Type A Uncertainty Evaluation}

\begin{figure}[!t]
	\centering		
	\includegraphics[width=2.6in]{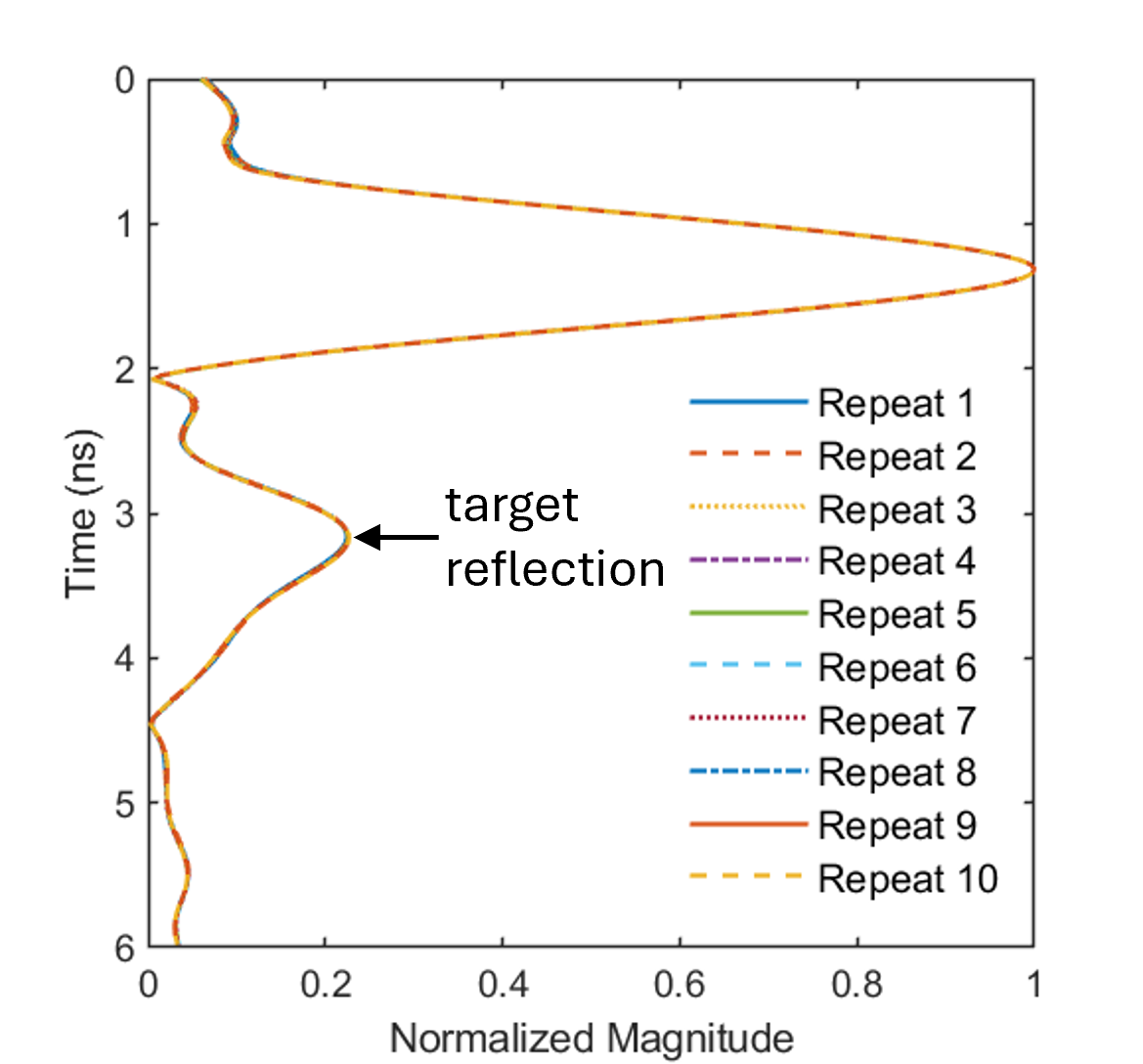}		
	\caption{Ten repeated center trace A-scans for a representative experimental configuration with a metallic rod placed along the $x$-axis and buried at a depth of $d = 16.4$~cm, an antenna-to-sand-surface distance of $d_{\mathrm{air}} = 20$~cm, and the antenna port axis is at $\phi=90^\circ$ relative to the rod axis.}
	\label{typeA}
\end{figure}

To assess the measurement repeatability, ten repeated B-scan measurements were performed using a representative experimental configuration. The metallic rod was buried at a depth of 16.40 cm, an antenna liftoff distance of 20 cm, and an antenna port orientation of $\phi = 90^{\circ}$ relative to the rod axis. The target position, scanning trajectory, antenna liftoff distance, and VNA settings were unchanged throughout the repeated B-scan measurements. Each measurement was independently processed using the same procedure described in Section \ref{sec:sectionIV}. For all ten repeated B-scans, the center A-scans corresponding to the antenna position directly above the target were extracted and shown in Fig.~\ref{typeA}. The ten responses exhibit close agreement throughout the entire time window, indicating high repeatability in measurement. \par

In accordance with the GUM Type A evaluation procedure \cite{JCGM2008}, the repeated measurements were statistically analyzed using the arithmetic mean, experimental standard deviation, and standard uncertainty of the mean. The target-reflection arrival time and amplitude were considered as two separate measurands. \par

For the target-reflection arrival time, the reflection peak occurred at 3.17383 ns for all ten repeated measurements. Therefore, the arithmetic mean was $\bar{t} = 3.17383$ ns, the experimental standard deviation was $s(t) = 0$ ns, and the Type A standard uncertainty of the mean was $u_A(\bar{t}) = 0$ ns, with 9 degrees of freedom. This indicates that no measurement-to-measurement variation in the target-reflection arrival time was resolved under the current temporal resolution. For the target-reflection amplitudes, the values obtained from the ten repeated measurements ranged from 0.224010 to 0.226802. The arithmetic mean was calculated as $\bar{A} = 0.225943$, and the experimental standard deviation was $s(A) = 8.35 \times 10^{-4}$. The Type A standard uncertainty of the mean was calculated as $u_A(\bar{A}) = 2.64 \times 10^{-4}$, with 9 degrees of freedom. Thus, the relative standard deviation of the repeated amplitude measurements was 0.37\%, indicating a high degree of repeatability.

\subsubsection{Type B Uncertainty Evaluation}

For the Type B uncertainty evaluation, uncertainty contributions not determined from the repeated measurements were considered. The main instrumentation-related source considered here is the CMT M5090 VNA \cite{m5090_datasheet}, which was used as the radar transceiver. \par

Using the target-reflection magnitude, the arithmetic mean of the ten measurements was $\bar{A}=0.225943$ on a linear scale, corresponding to approximately $-12.92$ dB. According to the manufacturer's specification, the M5090 has a calibrated reflection-magnitude accuracy of $\pm 0.4$ dB for reflection coefficients between $-15$ and 0 dB under the measurement conditions applicable to this experiment \cite{m5090_datasheet}. Given that the measured target-reflection magnitude of $-12.92$ dB lies within this range, the $\pm 0.4$ dB accuracy specification was used for the Type B evaluation. The manufacturer-specified accuracy was considered as a symmetric limiting interval. In the absence of available data specifying its probability distribution, a rectangular distribution was assumed following the GUM framework \cite{JCGM2008}. The corresponding Type B standard uncertainty is $u_B(A_{\mathrm{dB}})=0.4/\sqrt{3}=0.231$ dB.
Following the GUM law of propagation of uncertainty, the Type B standard uncertainty was then converted to the linear magnitude scale and calculated as $u_B(A)=[\ln(10)/20]\bar{A}u_B(A_{\mathrm{dB}})=[\ln(10)/20](0.225943)(0.231)=0.0060$. The corresponding relative Type B standard uncertainty is $u_{B,r}(A)=[u_B(A)/\bar{A}]\times100\%=(0.0060/0.225943)\times100\%=2.66\%$.\par

It is noted that the effects of robotic arm positioning repeatability and scanning path repeatability have been inherently captured in the ten repeated B-scan measurements in the Type A analysis, since each measurement involved a fully independent scanning cycle. These contributions are therefore not separately included in the Type B uncertainty to avoid double counting.

Overall, the uncertainty evaluation indicates a high degree of repeatability for the repeated measurements, with the Type A relative standard deviation of 0.37\%, and the Type B standard uncertainty associated with the VNA reflection-magnitude accuracy of 2.66\%. These results provide a quantitative assessment of the measurement repeatability and uncertainty of the proposed GPR system following the GUM framework.

\subsection{Comparison Between the UPR Antenna and the Single-Port Circularly Polarized (CP) Antenna}

The operating principles and detection characteristics of the proposed UPR antenna and a single-port circularly polarized (CP) antenna are compared in this section.
\subsubsection{Operating Principle}
A single-port CP antenna operates with one circular-polarization handedness, either left-hand circular polarization (LHCP) or right-hand circular polarization (RHCP). For a wave propagating in the $+z$ direction and following the IEEE convention, the normalized polarization vector for LHCP is
\begin{equation}
\hat{\rho}_{CP}
=
\frac{1}{\sqrt{2}}
\left(
\hat{x}+j\hat{y}
\right),
\label{eq:cp_pol}
\end{equation}
where it contains two orthogonal LP components of equal magnitude in phase quadrature. By comparison, the proposed UPR antenna is linearly polarized (LP) at each operating frequency, while its LP angle $\psi(f)$ changes continuously with frequency from approximately $0$ to $\pi$ over the 1.2--3.2 GHz operating band. Its normalized polarization vector is
\begin{equation}
\hat{\rho}_{\mathrm{UPR}}(f)
=
\cos\psi(f)\hat{x}
+
\sin\psi(f)\hat{y}.
\label{eq:upr_pol}
\end{equation}
As a result, the target is illuminated by a continuous range of LP angle within a single frequency sweep. 

\subsubsection{Detection of Elongated Targets}
For an idealized elongated target with its dominant axis oriented at an angle $\phi$, the target orientation can be written as
\begin{equation}
\hat{u}_{t}
=
\cos\phi\,\hat{x}
+
\sin\phi\,\hat{y}.
\label{eq:target_orientation}
\end{equation}
The effect of antenna polarization on the received radar response can be characterized by the polarization loss factor (PLF),
\begin{equation}
\mathrm{PLF}
=
\left(
\hat{\rho}\cdot\hat{u}_{t}
\right)^{2}.
\label{eq:plf}
\end{equation}
For a single-port CP antenna, $\mathrm{PLF}_{\mathrm{CP}}(f)=1/2$, which is independent of the target orientation. For the UPR antenna, $\mathrm{PLF}_{\mathrm{UPR}}(f)=\cos^{2}(\psi(f)-\phi)$. Since $\psi(f)$ varies continuously with frequency, different portions of the operating band correspond to different polarization alignments with the target. For an ideal UPR antenna with $\psi(f)$ varying linearly from $0$ to $\pi$ across the operating band, the average PLF over the full frequency range is $1/2$, independent of $\phi$. The ideal CP antenna has the same PLF of $1/2$. Therefore, from the polarization-alignment perspective, the ideal UPR and CP antennas are expected to provide comparable orientation-insensitive detection of elongated targets.

\subsubsection{Detection of Planar Targets}
The difference between the two antennas becomes more apparent when detecting predominantly specular planar targets. Taking an ideal planar PEC target as an example, the tangential electric-field components have identical reflection coefficients, $\Gamma_x=\Gamma_y=-1$. Under CP incidence, the reflected wave has the opposite CP handedness after reversal of the propagation direction, resulting in a PLF of 0 for the transmit and receive antennas with the same CP handedness. Consequently, the planar target cannot be effectively detected using the single-handedness CP antenna \cite{katz1963radar,nawc1997ewradar}. In contrast, for the UPR antenna, each frequency component is linearly polarized. Upon specular reflection from an ideal planar conductor, the LP field experiences a $180^{\circ}$ phase reversal while retaining its polarization axis. The reflected field therefore remains polarization matched with the UPR antenna at the corresponding frequency, allowing the planar-target response to be well received. \par

Overall, a single-port CP antenna can provide orientation-insensitive detection of elongated targets, but cannot well detect predominantly specular reflections from planar targets. In contrast, the proposed UPR antenna provides a range of LP orientations across frequency, enabling robust detection of both orientation-sensitive elongated targets and predominantly specular planar targets. Therefore, compared with both single-channel CP and single-LP antennas, the main advantage of the proposed UPR antenna is its ability to provide robust single-channel detection across targets with different geometries and orientations. \par

\subsection{Target-Orientation-Dependent Effective Bandwidth and Time-Domain Response}
As the LP angle of the UPR antenna varies continuously with frequency, the response of an elongated target depends on the frequency-dependent polarization alignment between the incident LP and the target orientation. Consequently, an elongated target exhibits a strong response only over the portion of the full antenna bandwidth where favorable polarization alignment occurs, resulting in reduced effective bandwidth of the target response, a broader A-scan pulse and reduced time-domain resolution.\par
To quantitatively demonstrate this effect, a representative case is considered in which the UPR antenna detects an elongated metallic target oriented at $\phi=90^\circ$. Over the 1.2–3.2 GHz operating frequency range, the LP angle $\psi(f)$ of the designed UPR antenna varies approximately from 0 to $\pi$. The frequency-dependent polarization alignment between the antenna polarization and the target orientation can be characterized by the PLF, calculated as $\mathrm{PLF}(f)=\cos^2\left(\psi(f)-\phi\right)$.\par
\begin{figure}[!t]
	\centering		
	\includegraphics[width=2.8in]{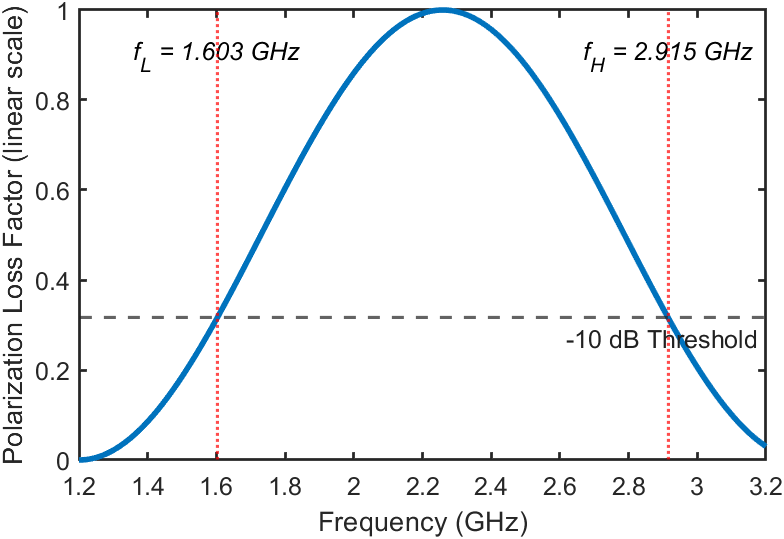}		
	\caption{Polarization loss factor as a function of frequency for a case in which the UPR antenna detects an elongated target oriented at $\phi=90^\circ$. The effective frequency range is defined using $-10$ dB threshold \cite{etsi_en302066_2020}.}
	\label{bw}
\end{figure}

The resulting PLF variation with frequency is shown in Fig.~\ref{bw}. Using $-10$ dB threshold on PLF to define the effective frequency \cite{etsi_en302066_2020}, the effective frequency range is from 1.603 GHz to 2.915 GHz, with an effective bandwidth of $B_{\mathrm{eff}}$ = 1.312 GHz. Compared with the full antenna bandwidth of 2.0 GHz, the effective bandwidth is 65.60\% of the full bandwidth. \par
The effect of this reduced effective bandwidth on the A-scan pulse width can be assessed from the inverse relationship between characteristic pulse width and bandwidth: $\Delta t \propto \frac{1}{B}$. Accordingly, the relative pulse-width change can be approximated as $\frac{\Delta t_{\mathrm{eff}}}{\Delta t_{\mathrm{full}}}=\frac{B_{\mathrm{full}}}{B_{\mathrm{eff}}}=1.5244$, corresponding to a 52.44\% increase in the pulse width. The time-domain range resolution in the demonstrated medium with relative permittivity $\varepsilon_r=3$ can be approximated as $\Delta z=\frac{c}{2\sqrt{\varepsilon_r}B}$. The range resolution changes from 4.33 cm for the full 2.0-GHz bandwidth to 6.60 cm for the 1.312-GHz effective bandwidth. \par

\section{Conclusion}
This paper presents a single-port ultrawideband polarization-rotating (UPR) antenna for improving detection robustness in GPR applications. The antenna consists of two orthogonally arranged Vivaldi radiators excited through a specifically designed feeding network that generates equal-amplitude signals with frequency-dependent phase differences, thereby enabling continuous linear polarization (LP) rotation from $0^\circ$ to $170^\circ$ over an ultrawide operating band of 1.2–3.2~GHz. Both simulated and measured results verify the polarization rotating behavior and broadband radiation performance of the antenna. \par
A single-channel stepped-frequency GPR system is then developed by integrating the UPR antenna, enabling rotating LP illumination of subsurface targets within a single measurement. Experimental results demonstrate that, compared with the conventional GPR system employing a fixed single LP antenna, the developed system provides more consistent detection performance across targets with different geometries and orientations. 
Enabled by its ability to achieve robust target detection without increasing hardware complexity, the developed antenna shows strong potential for many practical subsurface sensing applications.

\bibliographystyle{IEEEtran}
\bibliography{References}

\end{document}